\documentclass[a4paper,fleqn,usenatbib]{mnras}

\usepackage[T1]{fontenc}
\usepackage{ae,aecompl}

\usepackage{graphicx}	
\setkeys{Gin}{draft=false}
\usepackage{amsmath}	
\usepackage{amssymb}	
\usepackage{subfig}
\usepackage{threeparttable}

\usepackage{newtxtext,newtxmath}

\graphicspath{{Figures/}}

\def\MH2{M$_{\text{H}_2}$}
\def\arcsec{$^{\prime\prime}$}
\def\HI{H{\sc i}}
\newcommand{\xco}{X$_\text{CO}$}
\newcommand{\farc}{\mbox{\ensuremath{.\!\!^{\prime\prime}}}}
\newcommand{\kms}{\mbox{${\rm km}~{\rm s}^{-1}$}}

\newcommand\changes[1]{\textcolor{black}{#1}}
\newcommand\firstround[1]{\textcolor{black}{#1}}
\newcommand\secround[1]{\textcolor{black}{#1}}
\newcommand\refrep[1]{\textcolor{black}{#1}}

\title[AlFoCS + F3D II: low gas-to-dust ratios]{AlFoCS + F3D II: unexpectedly low gas-to-dust ratios in the Fornax galaxy cluster}

\author[N. Zabel et al.]{Nikki Zabel,$^{1}$\thanks{E-mail: zabel@astro.rug.nl}
Timothy A. Davis,$^{2}$
Matthew W. L. Smith,$^{2}$
Marc Sarzi,$^{3}$\newauthor
Alessandro Loni,$^{4, 5}$
Paolo Serra,$^{4}$
Maritza A. Lara-L\'opez,$^{3}$ 
Phil Cigan,$^{6}$ 
Maarten Baes,$^{7}$ \newauthor
George J. Bendo,$^{8}$
Ilse De Looze,$^{7, 9}$
Enrichetta Iodice,$^{10}$ 
Dane Kleiner,$^{4}$ \newauthor
B\"arbel S. Koribalski,$^{11, 12}$ 
Reynier Peletier,$^{1}$ 
Francesca Pinna,$^{13}$ 
P. Tim de Zeeuw$^{14, 15}$
\\
$^{1}$Kapteyn Astronomical Institute, University of Groningen, PO Box 800, 9700 AV Groningen, The Netherlands \\
$^{2}$School of Physics and Astronomy, Cardiff University, Queen's Building, The Parade, Cardiff, CF24 3AA, Wales, UK\\
$^{3}$Armagh Observatory and Planetarium, College Hill, Armagh, BT61 9DG, UK \\
$^{4}$INAF - Osservatorio Astronomico di Cagliari, Via della Scienza 5, I-09047 Selargius (CA), Italy\\
$^{5}$Dipartimento di Fisica, Universit\`a di Cagliari, Cittadella Universitaria, 09042, Monserrato, Italy \\
$^{6}$George Mason University, 4400 University Dr, Fairfax, VA 22030-4444, USA \\
$^{7}$Sterrenkundig Observatorium, Universiteit Gent, Krijgslaan 281 S9, B-9000 Gent, Belgium\\
$^{8}$UK ALMA Regional Centre Node, Jodrell Bank Centre for Astrophysics, School of Physics and Astronomy, The University of Manchester, \\ 
Oxford Road, Manchester M13 9PL, UK \\
$^{9}$Department of Physics and Astronomy, University College London, Gower Street, London WC1E 6BT, UK \\
$^{10}$INAF-Osservatorio Astronomico di Capodimonte, via Moiariello 16, 80131, Napoli, Italy \\
$^{11}$Australia Telescope National Facility, CSIRO Astronomy and Space Science, P.O. Box 76, 1710, Epping, NSW, Australia \\
$^{12}$Western Sydney University, Locked Bag 1797, 1797, Penrith South, NSW, Australia \\
$^{13}$Max-Planck-Institut f\"ur Aststuhl 17, 69117, Heidelberg, Germany \\
$^{14}$Sterrewacht Leiden, Leiden University, Postbus 9513, 2300 RA Leiden, The Netherlands \\
$^{15}$Max-Planck-Institut f\"ur extraterrestrische Physik, Giessenbachstra{\ss}e, 85741, Garching bei Muenchen, Germany \\
}

\date{Accepted 2021 February 3. Received 2021 February 1; in original form 2020 November 18}

\pubyear{2021}

\hypersetup{draft}

\begin{document}
\label{firstpage}
\pagerange{\pageref{firstpage}--\pageref{lastpage}}
\maketitle
\begin{abstract}
\firstround{We combine observations from ALMA, ATCA, MUSE, and \textit{Herschel} to study gas-to-dust ratios in 15 Fornax cluster galaxies detected in the \refrep{FIR/sub-mm} by \textit{Herschel} and observed by ALMA as part of the ALMA Fornax Cluster Survey (AlFoCS). The sample spans a stellar mass range of $8.3 \leq \text{\ log (M}_\star/\text{ M}_\odot) \leq 11.16$, and a variety of morphological types. We use gas-phase metallicities derived from MUSE observations (from the Fornax3D survey) to study these \firstround{ratios} as a function of metallicity, and to study dust-to-metal ratios, in a sub-sample of nine galaxies. We find that gas-to-dust ratios in Fornax galaxies are systematically lower than those in field galaxies at fixed stellar mass/metallicity. This implies that a relatively large fraction of the metals in these Fornax systems is locked up in dust, which is possibly due to altered chemical evolution as a result of the dense environment. The low ratios are not only driven by \HI\ deficiencies, but H$_2$-to-dust ratios are also significantly decreased. This is different in the Virgo cluster, where low gas-to-dust ratios inside the virial radius are driven by low \HI -to-dust ratios, while H$_2$-to-dust ratios are increased. \refrep{Resolved observations of NGC1436 show a radial increase in H$_2$-to-dust ratio, and show that low ratios are present throughout the disc.} We propose various explanations for the low H$_2$-to-dust ratios in the Fornax cluster, including the more efficient stripping of H$_2$ compared to dust, more efficient enrichment of dust in the star formation process, and altered ISM physics in the cluster environment.}
\end{abstract}




\section{Introduction}
\label{sec:introduction}
The environment in which a galaxy resides can significantly influence the way it evolves. In particular, the \refrep{more dense} the environment, the more early-type galaxies it harbours relative to late-type galaxies \citep{Oemler1974, Dressler1980}. This suggests that dense environments are capable of aiding the quenching of star formation in galaxies. Various environmental mechanisms have been proposed to be \mbox{(co-)responsible} for this, including ram pressure stripping \citep{Gunn1972}, starvation \citep{Larson1980}, and galaxy-galaxy interactions \citep{Moore1996}. While many of these processes primarily affect the more extended and easily stripped \HI\ gas, evidence has started to accumulate that they can also directly affect the \textit{molecular} gas, despite its more tightly bound and centrally located nature \citep[hereafter Z19]{Vollmer2008, Fumagalli2009, Boselli2014b, Zabel2019}. Since molecular gas is the direct fuel for star formation, this can, in turn, affect the star formation efficiency in galaxies in these dense environments (e.g. \citealt{Gullieuszik2020}, \citealt{Zabel2020}, hereafter Z20).

Besides gas, dust plays a vital role in star formation and galaxy evolution. If dust is environmentally stripped along with gas, this could have a significant impact on the evolution of galaxies in clusters, \secround{the most (galaxy )dense environments in the Universe.} Dust acts as a catalyst for the formation of H$_2$ \firstround{(e.g. \citealt{Dishoeck1993, Scoville2013, Wakelam2017} and references therein)}, and helps shield it from the interstellar radiation field \firstround{(e.g. \citealt{Dishoeck1993, Ciolek1995})}. \firstround{Therefore, dust and molecular gas are typically linked, and might be expected to be stripped simultaneously.} However, if the process of star formation is (indirectly) affected by dense environments, this may result in more subtle effects on the chemical evolution of a galaxy, and the relative gas and dust contents.

The gas-to-dust ratio provides a useful probe for studying the effects of environmental stripping on galaxies' gas and dust contents in more detail. Especially if it is compared with metallicity, it can teach us a lot about the chemical evolution of galaxies. The gas-to-dust ratio (G/D) as a function of metallicity has been studied extensively in field galaxies (e.g. \citealt{Issa1990, Lisenfeld1998, Draine2007a, Sandstrom2013, RemyRuyer2014, Vis2017b, Looze2020}). \citet{RemyRuyer2014} compiled and homogenised a range of recent samples. They found that \firstround{(total)} G/D correlates strongly with metallicity, following a broken (around 12 + log(O/H) = 7.96) power law, with large scatter ($\sim$0.37 dex for metallicity bins of $\sim$0.1 dex). Their results are consistent with chemical evolution models from \citet{Asano2013} and \citet{Zhukovska2016}. \secround{Resolved studies have shown that the gas-to-dust ratio also varies within individual galaxies, e.g. the local galaxy M33. Gas-to-dust ratios in M33 differ from that in the Milky Way, and also vary radially \citep{Tabatabaei2010, Gratier2017, Williams2019}.}

Studies of the Virgo cluster have shown that dust can be stripped by environmental effects as well as gas, even from the very inner parts of galaxies \citep{Cortese2010, Corbelli2012, Pappalardo2012}. \citet{Corbelli2012} find dust deficiencies of up to 75\% in \HI\ deficient Virgo galaxies. \citet{Pappalardo2012} find a strong spatial correlation between molecular gas and cold dust in non-\HI -deficient galaxies, which is weakened in \HI -deficient galaxies. This suggests that both molecular gas and dust are affected by environment, though not necessarily in the same way. They find no significant radial gradient in total \refrep{(i.e. \HI\ + H$_2$)} G/D within cluster galaxies. \firstround{\citet{Cortese2016} found that (total) gas-to-dust ratios in cluster galaxies are decreased compared to those in isolated galaxies at fixed mass and metallicity. They found that this decrease is driven by the \HI -to-dust ratio, while the H$_2$-to-dust ratio, on the other hand, is increased. They attribute this difference to differential stripping of the ISM as a result of the different spatial distribution of \HI\ (most extended), H$_2$ (centrally concentrated) and dust (in between the two gas phases).}

Thus far, studies of gas-to-dust ratios in cluster galaxies have mostly focused on spiral galaxies in the Virgo cluster. To obtain a broader understanding of the role of dust stripping in the baryon cycle of cluster galaxies, in a wider variety of environments, we need to expand our studies to different clusters and galaxy types. The Fornax cluster is the more mature \secround{(though still active, e.g. Z19; \citealt{Iodice2019b, Raj2020, Loni2021})}, but smaller sibling of the Virgo cluster in the Southern sky. Although it has only $\sim$1/10 times the mass of Virgo ($\sim7 \times 10^{13} M_\odot$ within twice the virial radius, \refrep{2 $\times$ 0.7 = 1.4 Mpc, the virial radius of the Virgo cluster is $\sim$1.65 Mpc}, \citealt{Drinkwater2001a, Mamon2004, Jordan2007}), and $\sim$1/6 times the number of galaxies \citep{Binggeli1985, Ferguson1989}, its galaxy number density is 2-3 times higher. It is also more \firstround{homogeneous} and dynamically evolved, which is probably more typical of the type of environment many galaxies in the local universe reside in. Located at a distance similar to that to Virgo \refrep{(the distance to the Fornax cluster is 19.95 Mpc, while that to the Virgo cluster is 16.5 Mpc \citealt{Tonry2001, Mei2007})}, the Fornax cluster provides an excellent second laboratory for studying environmental processes in detail.

The ALMA Fornax Cluster Survey \firstround{(AlFoCS)} is a complete survey of galaxies in the Fornax cluster that were detected in the far infrared with \textit{Herschel} or in \HI, with the goal to study the effects of environment on molecular gas in combination with other phases of the interstellar medium (ISM). It comprises 30 galaxies \refrep{(of which 15 were detected in CO)} of various morphological types, both early-types and late-types, as well as (irregular) dwarf galaxies, spanning a mass-range of $10^{\sim 8.5 - 11}$ M$_\odot$. \refrep{for reference, the optical Fornax Cluster Catalogue (FCC, \citealt{Ferguson1989}) contains 340 galaxies, and with the recent optical Fornax Deep Survey (FDS, \citealt{Peletier2020}), this number probably increases to several thousands of galaxies, with many new photometric cluster candidates (564 dwarf galaxies alone) that have been identified in \citet{Venhola2018}}. In Z19 we presented resolved molecular gas maps and H$_2$ masses and deficiencies. In Z20, \refrep{where we use both Atacama Large Millimiter/submillimeter Array (ALMA) and Multi Unit Spectroscopic Explorer (MUSE) observations (see below)}, we studied the resolved star formation relation in the Fornax cluster. In this work, we study molecular gas-to-dust ratios in relation to their metallicities.

\firstround{Recently, a $15 \times 15$ degree blind survey, covering the Fornax cluster out to its virial radius, was performed with the Australia Telescope Compact Array (ATCA, \citealt{Loni2021}). This survey resulted in the detection of 16 Fornax galaxies in \HI. 15 of these have CO observations or useful upper limits \refrep{(i.e. upper limits that provide informative constraints on the H$_2$-to-dust ratio)} from AlFoCS, allowing us to study their total gas-to-dust ratios.}

Fornax3D (F3D, \citealt{Sarzi2018, Iodice2019b}) targeted all galaxies from the Fornax Cluster Catalogue \citep{Ferguson1989} brighter than $m_B = 15$ within or close to the virial radius ($R_\text{vir}$ = 0.7 Mpc; \citealt{Drinkwater2001a}) with MUSE, mounted to the Yepun Unit Telescope 4 at the Very Large Telescope (VLT). \firstround{Nine of these galaxies were detected (or have a useful upper limit) with AlFoCS and ATCA.}

This paper is organised as follows. In \S \ref{sec:observations} we describe our sample and observations. In \S \ref{sec:methods} we describe the methods used, in particular how any masses and metallicities were estimated. \firstround{In \S \ref{sec:lit_samples} we describe the DustPedia literature sample, to which we compare our results, and any underlying assumptions used to derive the quantities we utilise. In \S \ref{sec:results} we describe the main results. In \S \ref{sub:NGC1436} we look into the resolved H$_2$-to-dust ratio in NGC1436, to highlight any radial or other spatial variation, if present. The main results are discussed in \S \ref{sec:discussion}}. Finally, we summarise our findings in \S \ref{sec:summary}. 

Throughout this work, we assume a common distance of 19.95 Mpc \citep{Tonry2001} to galaxies in the Fornax cluster.

\section{Sample selection, observations and data reduction}
\label{sec:observations}

\subsection{The sample}
\label{sub:sample_fornax}
Our Fornax sample consists of all AlFoCS galaxies that have \refrep{far-infrared/sub-millimetre} measurements from which dust masses can be estimated reliably (i.e. are detected in $\geq$3 \textit{Herschel} bands, \refrep{implying at least one detection with the Spectral and Photometric Imaging REceiver (SPIRE, \citealt{Griffin2010}, see \S \ref{sub:dust_data_fornax})}. It consists of 15 galaxies, spanning a mass range of $8.3 \leq$ log(M$_\star$/M$_\odot$) $\leq 11.16$. Both late-type and early-type galaxies are included in the sample, as well as (irregular) dwarfs. Some of the \firstround{H$_2$} measurements are upper limits, though these are only included if they provide a useful constraint on the H$_2$-to-dust ratio. Of these galaxies, \firstround{9 (of which one \secround{having a} CO upper limit)} have MUSE observations from F3D, and can therefore be used to study H$_2$-to-dust ratios as a function of metallicity. This sub-sample spans a stellar mass range of $8.3 \leq$ log(M$_\star$/M$_\odot$) $\leq 11.0$. Key parameters of the sample are listed in Table \ref{tab:targets_fornax}. Fornax Cluster Catalogue (FCC, \citealt{Ferguson1989}) numbers are listed in column one. Corresponding common galaxy names are listed in column 2. \refrep{Column 3 lists stellar masses, from z0mgs \citep{Leroy2019}, or from \citet{Fuller2014} if not available there, indicated with a $\ddagger$.} Morphological types are given in column 4, from \citet{Sarzi2018} where available, otherwise from older references provided in the Table caption. Column 5 lists the projected cluster-centric distance in kpc. Column 6 describes whether the molecular gas reservoir is regular (R) or disturbed (D), as classified in Z19. Column 7 lists \xco, estimated as described in \S \ref{sub:H2_masses}. Molecular gas, atomic gas, and dust masses are listed in columns 8, 9, and 10, respectively. Finally, column 11 lists whether the galaxy is included in F3D (Y) or not (N).

\begin{table*}
	\centering
	\begin{threeparttable}
	\caption{Key properties of the galaxies in the sample.}
	\label{tab:targets_fornax}
	\setlength{\tabcolsep}{1.5mm}
	\begin{tabular}{rrrrrrrrrrrr}
	\hline
	FCC & Common name & M$_\star$ & Type & D & Gas dist. & \xco & M$_{\text{H}_2}$ & M$_{\text{H}{\textsc i}}$ & M$_\text{dust}$ & in F3D? \\
	- & - & (log $M_{\odot}$) & - & (kpc) & - & $10^{20}$ cm$^{-2}$ (K km s$^{-1}$)$^{-1}$ & (log $M_\odot$) & (log $M_\odot$) & (log $M_\odot$) & (Y/N) \\
	(1) & (2) & (3) & (4) & (5) & (6) & (7) & (8) & (9) & (10) & (11) \\
	\hline
	67 & NGC1351A & 9.56 $\pm$ 0.1 & SB(rs)bc\textsuperscript{a} & 694 & R & 2.2 $\pm$ 0.7 & 7.03 $\pm$ 0.07 & 8.67 $\pm$ 0.11 & 6.637$^{+0.051}_{-0.052}$ & N \\
	90 & MGC-06-08-024 & 8.98$^\ddagger$ & E4 pec & 595 & D & 6.6 $\pm$ 0.7 & 7.33 $\pm$ 0.07 & 7.76 $\pm$ 0.16 & 5.234$^{+0.117}_{-0.114}$ & Y \\
	121 & NGC1365 & 10.75 $\pm$ 0.1 & SB(s)b\textsuperscript{b} & 420 & R & 1.6 $\pm$ 0.7 & 9.49 $\pm$ 0.04 & 10.18 $\pm$ 0.09 & 8.093$^{+0.029}_{-0.029}$ & N \\
	167 & NGC1380 & 8.75 $\pm$ 0.11 & S0/a & 220 & R & 0.9 $\pm$ 0.7 & 7.43 $\pm$ 0.06 & $\leq$7.7 & 5.873$^{+0.039}_{-0.039}$ & Y \\
	179 & NGC1386 & 10.92 $\pm$0.09 & Sa & 226 & R & 1.3 $\pm$ 0.7 & 8.25 $\pm$ 0.04 & $\leq$7.7 & 6.569$^{+0.031}_{-0.032}$ & Y \\	
	184 & NGC1387 & 9.51 $\pm$ 0.18 & SB0 & 111 & R & 1.1 $\pm$ 0.7 & 8.14 $\pm$ 0.04 & $\leq$7.5 & 6.359$^{+0.032}_{-0.031}$ & Y \\
	235 & NGC1427A & 9.1 $\pm$ 0.17 & IB(s)m\textsuperscript{b} & 132 & - & 2.6 $\pm$ 0.7 & $\leq$7.42 & 9.32 $\pm$ 0.09 & 6.446$^{+0.113}_{-0.120}$ & N \\
	263 & PGC013571 & 8.75 $\pm$ 0.1 & SBcdIII & 292 & D & 11.1 $\pm$ 0.7 & 7.88 $\pm$ 0.05 & 8.01 $\pm$ 0.13 & 5.938$^{+0.062}_{-0.062}$ & Y \\
	282 & FCC282 & 8.86 $\pm$ 0.1 & S0\textsuperscript{c} & 615 & D & 2.8 $\pm$ 0.7 & 7.15 $\pm$ 0.05 & $\leq$7.3 & 5.362$^{+0.085}_{-0.084}$ & N \\
	285 & NGC1437A & 8.71 $\pm$ 0.1 & SAB(s)dm\textsuperscript{b} & 428 & - & 3.7 $\pm$ 0.7 & $\leq$7.83 & 8.75 $\pm$ 0.10 & 6.508$^{+0.212}_{-0.199}$ & Y \\ 
	290 & NGC1436 & 10.12 $\pm$ 0.1 & ScII & 389 & R & 1.2 $\pm$ 0.7 &8.28 $\pm$ 0.05 & 8.21 $\pm$ 0.19 & 6.817$^{+0.037}_{-0.036}$ & Y \\
	306 & FCC306 & 8.68$^\ddagger$ & S\textsuperscript{c} & 600 & - & 3.9 $\pm$ 0.7 & $\leq$8.10 & 8.03 $\pm$ 0.13 & 4.478$^{+0.322}_{-0.319}$ & N \\
	308 & NGC1437B & 9.25 $\pm$ 0.11 & Sd & 611 & D & 5.8 $\pm$ 0.7 & 8.17 $\pm$ 0.04 & 8.39 $\pm$ 0.11 & 6.597$^{+0.064}_{-0.068}$ & Y \\
	312 & ESO358-G063 & 10.08 $\pm$ 0.11 & Scd & 584 & R & 4.7 $\pm$ 0.7 & 8.99 $\pm$ 0.05 & 9.23 $\pm$ 0.10 & 7.057$^{+0.029}_{-0.030}$ & Y \\
	335 & ESO359-G002 & 9.17 $\pm$ 0.1 & SB0\^-?\textsuperscript{b} & 872 & D & 2.4 $\pm$ 0.7 & 6.92 $\pm$ 0.05 & $\leq$7.0 & 4.787$^{+0.101}_{-0.100}$ & N \\
	\hline
	\end{tabular}
	\textit{Notes:} \textbf{1}: Fornax Cluster Catalogue \citep{Ferguson1989} number of the galaxy; \textbf{2}: Common name of the galaxy; \textbf{3}: \refrep{Stellar mass from z0mgs \citep{Leroy2019}, or from \citet{Fuller2014} if not available there, indicated with a $\ddagger$}; \textbf{4}: Morphological type from \citet{Sarzi2018} if available, otherwise references are listed below; \textbf{5}: Projected distance from brightest cluster galaxy NGC1399; \textbf{6}: Whether the molecular gas in the galaxy is regular (R) or disturbed (D) as classified in Z19; \textbf{7}: \xco, estimated as described in \S \ref{sub:H2_masses}; \textbf{8}: H$_2$ mass from Z19; \textbf{9}: \HI\ mass from \citet{Loni2021}; \textbf{10}: Dust mass calculated from the FIR fluxes published in \citet{Fuller2016}; \textbf{11}: Whether the galaxy was observed with MUSE as part of F3D (Y) or not (N).
	
	\textsuperscript{a} \citep{Lauberts1989}; \textsuperscript{b} \citep{Vaucouleurs1991}; \textsuperscript{c} \citep{Loveday1996}
	\end{threeparttable}
\end{table*}

\subsection{CO data}
\label{sub:co_data_fornax}
\firstround{ALMA data for our Fornax cluster targets were analysed in Z19, which describes the data and methods used in detail.} We summarise some important details here. ALMA Band 3 observations were carried out between the 7th and 12th of January 2016 under project ID 2015.1.00497.S (PI: T. Davis), using the main (12m) array in the C36-1 configuration. The data were calibrated manually, \textsc{clean}-ed interactively, using a natural weighting scheme, and continuum subtracted using the Common Astronomy Software Applications package (CASA, version 5.1.1, \citealt{McMullin2007}). \firstround{The FWHM of the restoring beam is typically between $\sim$2$^{\prime\prime}$ and 3$^{\prime\prime}$ (equivalent to $\sim$200 - 300 pc at the distance of the Fornax cluster).} Channel widths are 10 \kms. Typical rms noise levels are $\sim$3 mJy/beam. The cleaned data cubes were used to produce moment maps of the $^{12}$CO(1-0) line emission using the masked moment method from \citet{Dame2011}. \refrep{These maps were corrected for the primary beam response.} Spectra, from which the H$_2$ masses were estimated, were calculated by summing over both spatial directions of the spectral cube, using a square spatial field around the emission. \refrep{At the distance of the Fornax cluster, the largest scales recoverable with the 12m array are significantly larger than the expected sizes of the largest CO structures the galaxies observed. Therefore, we expect to have recovered the total CO(1-0) flux of each galaxy, and we expect the masses derived in Z19 to be accurate.} For one object, NGC1365, ALMA data was added from the archive (project ID: 2015.1.01135.S, PI: Fumi Egusa, see Z19 for more details). \refrep{For a more detailed, resolved study of H$_2$-to-dust ratios in NGC1436, described in \S \ref{sub:NGC1436}, we use additional, deeper ALMA data from the archive to complement the observations from Z19 (project ID: 2017.1.00129.S, PI: Kana Morokuma, see \S \ref{sub:NGC1436} for more details).}

\subsection{Far-infrared data}
\label{sub:dust_data_fornax}
The far-infrared (FIR) maps used to estimate dust masses are from the Photoconductor Array Camera and Spectrometer (PACS, \citealt{Poglitsch2010}, 100 and 160 micron) and the Spectral and Photometric Imaging REceiver (SPIRE, \citealt{Griffin2010}, 250, 350, and 500 micron), both mounted on the \textit{Herschel} Space Observatory \citep{Pilbratt2010}. \changes{The SPIRE maps used here are identical to the ones used by DustPedia (\citealt{Davies2017}, see \S \ref{sec:lit_samples}), and the PACS maps were reprocessed using the same techniques as DustPedia.} \secround{The FWHM of the beams of the \textit{Herschel} maps are $\sim 6.8$, $\sim 10.7$, $\sim 17.6$, $\sim 23.9$, $\sim 35.2^{\prime\prime}$ at 100, 160, 250, 350, and 500 micron, respectively, corresponding to $\sim 660, \sim 1000, \sim 1700, \sim 2300,$ and $\sim 3400$ pc at the distance of the Fornax cluster \refrep{(see the SPIRE handbook\footnote{\label{foot:spire_handbook}\url{http://herschel.esac.esa.int/Docs/SPIRE/spire_handbook.pdf}} and the PACS Observer's Manual\footnote{\url{https://www.cosmos.esa.int/documents/12133/996891/PACS+Observers\%27+Manual}}).}}

Note that FIR measurements and derived dust masses for the Fornax cluster are already available in \citet{Fuller2014}, on which the AlFoCS sample is based. However, since this work was published, several improvements have been made both to the \textit{Herschel} data reduction and the SED fitting methods. In order to obtain dust masses that are as accurate as possible, we have opted to re-estimate dust masses from improved FIR maps, using updated SED fitting techniques (see \S \ref{sub:dust_masses}). \firstround{Dust masses calculated here are slightly lower than those published in \citet{Fuller2014}, with a median difference of 0.28 dex and a 1$\sigma$ spread in differences of 0.28 dex.}

\subsection{\HI\ data}
\label{sub:hi_data}
\firstround{\HI\ data\footnote{Available from \url{https://atoa.atnf.csiro.au/query.jsp}} are from ATCA, which was used to conduct a blind survey of the Fornax cluster, covering an area of 15 deg\textsuperscript{2} out to $R_\text{vir}$. The observations and data reduction are presented and described in detail in \citet{Loni2021}, and are summarised here. Observations were carried out from December 2013 to January 2014 in the 750B configuration (project code: C2894, PI: P. Serra). The data were reduced manually using the \texttt{MIRIAD} software \citep{Sault1995}. The dirty cube was obtained using the \texttt{INVERT} task \secround{(using natural weighting)}, after which the tasks \texttt{MOSMEM} and \texttt{RESTOR} were used to clean and restore the \HI\ emission. The synthesised beam has a FWHM of $95^{\prime\prime} \times 67^{\prime\prime}$ (corresponding to $\sim 9 \times 6.5$ kpc at the distance of the Fornax cluster). Channel widths are 6.6 \kms . Typical noise levels are 2.8 mJy beam$^{-1}$ and go down to 2.0 mJy beam$^{-1}$ in the most sensitive region. \HI\ sources were identified using the SoFiA source-finding package \citep{Serra2015}. Whether detections are considered reliable is based on the algorithm from \citet{Serra2012}, and by visual inspection where necessary. This has resulted in the detection of \HI\ in 16 Fornax cluster galaxies, of which 15 have CO detections (or useful upper limits) and are thus included in this work. \refrep{The spatial resolution of the ATCA data is 67$^{\prime\prime}$ $\times$ 95$^{\prime\prime}$ ($\sim$ 6 $\times$ 9 kpc at the distance of the Fornax cluster, see \citealt{Loni2021}). This means that the \HI\ discs are marginally resolved in this data set.}}
 
\subsection{Optical spectra}
\label{sub:optical_spectra_fornax}
Optical spectra, used to measure the line ratios from which metallicities are estimated, are from F3D \cite{Sarzi2018, Iodice2019b}. A detailed description of the survey and data reduction can be found in \citet{Sarzi2018}, and some important details are summarised here. 

Integral-field spectroscopic observations were carried out with MUSE (\citealt{Bacon2010}, mounted to VLT Unit Telescope 4, ``Yepun'') in Wide Field Mode, between July 2016 and December 2017. A field of 1 $\times$ 1 square arcminutes was covered, with 0.2 $\times$ 0.2 square arcsecond spatial sampling. For some of the more extended galaxies, this is smaller than their optical discs. In these cases two or three pointings were used to map the whole galaxy. An exception is FCC290, which was only observed partially (including the centre and the outskirts \firstround{to one side, see \citealt{Sarzi2018}}). The MUSE pointings of all F3D galaxies can be found in \citet{Sarzi2018}. The observations cover a wavelength range of 4650-9300 \AA, with a spectral resolution of 2.5 \AA\ at the full width at half maximum (FWHM) at 7000 \AA\ and spectral sampling of 1.25 \AA\ pixel$^{-1}$. 

Data reduction was performed using the MUSE pipeline (version 1.6.2, \citealt{Weilbacher2012, Weilbacher2016}) under the ESOREFLEX environment \citep{Freudling2013}. In summary, the data reduction involved bias and overscan subtraction, flat fielding, wavelength calibration, determination of the line spread function, illumination correction with twilight flats (to account for large-scale variation of the illumination of the detectors) and similar with lamp flats (to correct for edge effects between the integral-field units). 

\section{Methods}
\label{sec:methods}

\subsection{H$_2$ masses}
\label{sub:H2_masses}
H$_2$ masses of the Fornax galaxies are calculated as described in \S 4.3 of Z19. \firstround{Upper limits are one beam, 3$\sigma$ upper limits, assuming a linewidth of 50 \kms, as described in \S 4.3.1 of Z19.} In summary, we use the following equation:
\begin{equation}
\label{eq:MH2}
M_{\text{H}_2} = 2 m_\text{H}\ D^2\ X_\text{CO}\ \frac{\lambda^2}{2 k_\text{B}}\ \int S_\nu \text{d}v,
\end{equation}
where $m_\text{H}$ is the mass of a hydrogen atom, $D$ the distance to the galaxy (in this case assumed to be the distance to the Fornax cluster), $X_\text{CO}$ is the CO-to-H$_2$ mass conversion factor, $\lambda$ the rest wavelength of the line observed, $k_\text{B}$ the Boltzmann constant, and $\int S_\nu \text{d}v$ the total flux of the line observed. For $X_\text{CO}$ we use the metallicity-dependent mass conversion factor $\alpha_\text{CO}$ from \citet[eqn. 25]{Accurso2017}:
\begin{equation}
\label{eq:XCO}
\text{log}\ \alpha_\text{CO} = 14.752 - 1.623 \left[ 12 + \text{log(O/H)} \right] + 0.062\ \text{log}\ \Delta \text{(MS)},
\end{equation}
where $12 + \text{log(O/H)}$ is the metallicity as estimated according to \S \ref{sub:metallicities}, and $\Delta \text{(MS)}$ the distance from the main sequence (from \citealt{Elbaz2007}) as estimated in Z20. $\alpha_\text{CO}$ is then converted to $X_\text{CO}$ by multiplying it by $2.14 \times 10^{20}$. Note that this prescription includes a correction for helium, however for convenience we will use $M_{\text{H}_2}$ to refer to the total molecular gas mass. The results are listed in Table \ref{tab:targets_fornax}.

\subsection{\HI\ masses}
\label{sub:HI_data}
\HI\ masses are from \citet{Loni2021}, who use the prescription from \citet{Meyer2017} to convert integrated fluxes to \HI\ masses. \firstround{A common distance of 20 Mpc was assumed for all galaxies, a negligable difference with the 19.95 Mpc assumed in this work.} In cases where there is no detectable \HI, upper limits are calculated. These are estimated as 3 $\times$ the local rms noise in the ATCA data cube, in one beam, and assuming the linewidths of the corresponding CO(1-0) lines from Z19. The resulting \HI\ masses are summarised in Table \ref{tab:targets_fornax}.

\subsection{Dust masses}
\label{sub:dust_masses}
Dust masses are estimated directly from the \textit{Herschel} maps described in \S \ref{sub:dust_data_fornax}. To obtain FIR flux measurements for each galaxy, we perform aperture photometry, using the \textsc{Python} package \textsc{photutils} \citep{Bradley2019}. \firstround{Reasonable aperture extents were determined by eye, such that they encompass the dust emission from the entire galaxy} \secround{(typical semimajor axis-sizes vary from 15$^{\prime\prime}$ for small galaxies to 400$^{\prime\prime}$ for NGC1365, which is slightly smaller than the extent of the \HI\ disc in that galaxy, and closer to the extent of its stellar disc)}. To ensure all flux is included even in the lowest-resolution image, we define apertures in the longest wavelength image in which the galaxy is detected. In case there is no detection in each band, upper limits from the non-detected bands are used to constrain the SED at those wavelengths, which is reflected in the uncertainties. Background noise was estimated locally, by defining annuli around the apertures chosen. The flux in these annuli is then subtracted from the measured source flux. Uncertainties were estimated by randomly placing 100 apertures of the same dimensions as the one used for the source randomly in the \textit{Herschel} map, and calculating the standard deviation in these. This is done for each source in each of the 5 wavelengths. \firstround{The maps used are parallel-mode scan maps that cover very large areas (roughly 4.6 $\times$ 4.6 degrees, corresponding to $\sim 1.6 \times 1.6$ Mpc at the distance of the Fornax cluster). This means that it is relatively easy to find 100 apertures that do not overlap with each other, and there is only a small probability of these random apertures overlapping with other sources.} However, to exclude any such overlap from the noise estimate, the resulting flux distributions are clipped beyond 3 $\sigma$ from the median.

For small sources, the SPIRE beam is often larger than the aperture used to measure the flux: the full width at half maximum (FWHM) is up to $\sim$35$^{\prime\prime}$, depending on the wavelength (see the SPIRE handbook\refrep{\textsuperscript{\ref{foot:spire_handbook}}}, while the radii of our apertures are as small as 15$^{\prime\prime}$ for the smallest sources. To account for any missing flux as a result of this, \firstround{which might result in an underestimation of dust masses (and therefore an overestimation of gas-to-dust ratios),} we apply aperture corrections. As the sources are not very extended at \textit{Herschel} resolutions, we apply the recommended corrections for point-sources, provided as part of the SPIRE calibration \citep{Ott2010, Bendo2013}. These values depend on the aperture radius, which we estimate by taking the \firstround{arithmetic} average of the semi-major and semi-minor axes of the elliptical apertures.

Finally, dust masses are estimated from the resulting SEDs. We perform modified blackbody fits, described by \firstround{(a simplified version of) equation 1 in \citet{Hildebrand1983}:}
\begin{equation}
F_\lambda = \kappa_\lambda\ B_\lambda \left(T\right) D^{-2} M_\text{dust},
\end{equation}
\changes{where $M_\text{dust}$ the dust mass in kg, $D$ the distance to the galaxy in Mpc, $B$ (T) Plank's law of blackbody radiation, and $\kappa_\lambda$ is described by}
\begin{equation}
\kappa_\lambda = \kappa_0 \left( \lambda_0 / \lambda \right) ^\beta,
\end{equation}
where $\beta$ is the dust opacity index, which we fix at \firstround{1.790 to match our comparison sample (DustPedia, see \S \ref{sec:lit_samples})}, and $\kappa$ is the \changes{dust emissivity}, which we fix at 0.192 m$^2$ kg$^{-1}$ at $\lambda$ = 350 $\mu$m, again following the value used by DustPedia. \firstround{Both constants adopted by DustPedia are drawn from The Heterogeneous dust Evolution Model for Interstellar Solids (THEMIS) model \citep{Jones2013, Koehler2014, Ysard2015}.} The fits were performed using our own SED fitter, which is based on \textsc{PyMC3} (a Python package for Bayesian statistical modelling and Probabilistic Machine Learning focusing on advanced Markov chain Monte Carlo (MCMC) and variational inference (VI) algorithms, \citealt{Salvatier2016}). It was written to include all beam and colour corrections as part of the fitting process, and account for correlated uncertainties between bands. 

\refrep{For the (log) dust mass, we use a Gamma prior with a standard deviation of 1 dex. To ensure that the code can handle a wide range of dust masses (from a small region of a galaxy to a ULIRG), the mode is programmed to be the flux at the data point closest to the peak of the blackbody. For dust temperature, we also use a Gamma distribution, with a mode of 20K and a standard deviation of 6K. We choose the Gamma distribution as it does not allow for unphysical negative dust temperatures. It is also wider than a Gaussian distribution, which more accurately reflects the wide range of dust mass and temperatures.}

The assumed calibration uncertainties are 5\% for PACS \citep{Balog2014}, and 1.7\% for SPIRE \citep{Bendo2013}, as well as an additional correlated uncertainty of 5\% between PACS bands and 4\% between SPIRE bands. \firstround{The correlated and uncorrelated uncertainties were conservatively added linearly for the SPIRE data as recommended by the handbook\textsuperscript{\ref{foot:spire_handbook}}, resulting in a total uncertainty of 5.5\%. For an overview of the general limitations of SED fitting techniques (in particular, dust emission in the FIR appearing to be a blend of emission from dust at different temperatures), we refer the reader to \citet{Bendo2015}.}

\subsection{Metallicities}
\label{sub:metallicities}
\changes{Throughout this work, we use the oxygen abundance, 12 + log(O/H), \firstround{as a proxy for} the gas-phase metallicity.} Metallicities are estimated from F3D emission-line measurements presented in \cite{Iodice2019b}, and using the strong-line calibration from \citet[][referred to as DOP16 throughout the rest of this work]{Dopita2016}. This calibration relies exclusively on the [N\textsc{II}]$\lambda$6484, S[\textsc{II}]$\lambda\lambda$6717, 31, and H$\alpha$ lines, as follows: 
\begin{equation}
12 + \text{log(O/H)} = 8.77 + \text{log[N\textsc{II}]/[S\textsc{II}]} + 0.264 \text{ log[N\textsc{II}]/H}\alpha.
\end{equation}
\firstround{It is independent of the ionisation parameter, flux calibration, and extinction correction, and valid for a wide range of oxygen abundances.}
The required emission-line fluxes were estimated by simultaneously fitting the spectral continuum and the nebular emission lines using Gas AND  Absorption Line Fitting (\textsc{gandalf}, \citealt{Sarzi2006}), which \firstround{makes use of Penalized Pixel-Fitting (\texttt{PPXF}, \citealt{Cappellari2017}) to derive stellar kinematic and corresponding absorption-line broadening}. More details on this procedure can be found in \citet{Sarzi2018} and \citet{Iodice2019b}. The resulting resolved metallicity measurements were averaged spatially to obtain global measurements. 

\refrep{Of course, the choice of metallicity calibrator can significantly impact the resulting metallicity estimates. As can be seen in e.g. \citet{Sanchez2019}, who compare various metallicity calibrations as a function of stellar mass, and SFR, the DOP16 calibration shows a relatively steep gradient with stellar mass. It agrees reasonably well with the calibration from PP04 at the low-mass end, with an offset of $\sim$0.1 dex, while at the higher-mass end it over predicts metallicities by up to $\sim$0.4 dex compared to PP04. A similar study can be found in Appendix C of \citet{Kreckel2019}. As a result, the use of a different metallicity calibrator would likely result in the higher metallicities becoming somewhat lower, and the opposite for the lowest metallicities. As we will see in \S \ref{sec:results}, a different choice of metallicity calibrator (i.e. shifting the data along the x-axes of the figures showing gas-to-dust ratios as a function of metallicity, making the distribution more compact) would not affect our results. More importantly, the DOP16 calibrator was applied to both the Fornax sample and the comparison sample, which means that the offset in gas-to-dust ratio as a function of metallicity between the two would not be different had we chosen another calibrator.}

\subsection{Stellar masses}
\label{sub:stellar_masses}
\refrep{Stellar masses are from the z = 0 Multiwavelength Galaxy Synthesis (z0MGS, \citealt{Leroy2019}) where available. These are based on data from the Wide-field Infrared Explorer (WISE, \citealt{Wright2010}) and the Galaxy Evolution Explorer (GALEX, \citealt{Martin2005}). \citet{Leroy2019} make use of data from the GALEX-SDSS-WISE Legacy Catalog (GSWLC, \citealt{Salim2016, Salim2018}), which combines WISE, GALEX, and Sloan Digital Sky Survey (SDSS) observations. In the GSWLC SED modeling with the Code Investigating GALaxy Emission (CIGALE, \citealt{Burgarella2005, Noll2009, Boquien2019}) was used to yield stellar mass estimates. In \citet{Leroy2019} these data are used to derive calibrations to estimate stellar masses from their sample of WISE and GALEX data. The IMF from \citet{Kroupa2003} was assumed.}

\refrep{If stellar masses are not available from z0MGS (this only applies to two dwarf galaxies, FCC090 and FCC306), they are taken from \citet{Fuller2014}. Although the stellar masses in \citet{Fuller2014} are estimated differently from those in z0MGS, even a significant error in the stellar masses of these two dwarf galaxies would not affect our results. Stellar masses, along with their sources, are listed in Table \ref{tab:targets_fornax}. The two stellar masses that were taken from \citet{Fuller2014} are indicated with a $\ddagger$.}


\section{The DustPedia comparison sample}
\label{sec:lit_samples}

\begin{table*}
	\centering
	\begin{threeparttable}
	\caption{Estimated gas-to-dust ratios and residuals of galaxies in the sample.}
	\label{tab:gas-to-dust_ratios}
	\setlength{\tabcolsep}{1.5mm}
	\begin{tabular}{rrrrrrrrrr}
	\hline
 Object & Total gas/dust & res. ($\text{M}_\star$) & res. (Z) & M$_{\text{H}_2}$/dust & res. ($M_\star$) & res. (Z) & M$_{\text{H}\textsc{i}}$/dust & res. ($M_\star$) & res. (Z) \\- & - & (dex) & (dex) & - & (dex) & (dex) & - & (dex) & (dex) \\ (1) & (2) & (3) & (4) & (5) & (6) & (7) & (8) & (9) & (10) \\ \hline 
 FCC067 & 110 $^{+ 20 } _{- 20 }$ &-0.75& - & 2 $^{+ 0 } _{- 0 }$ &-1.97& - & 100 $^{+ 10 } _{- 10 }$ &-0.67& - \\
 FCC090 & 460 $^{+ 150 } _{- 140 }$ &-0.13&0.02& 130 $^{+ 20 } _{- 20 }$ &-0.26&-0.06& 340 $^{+ 90 } _{- 90 }$ &-0.18& -0.07 \\
 FCC121 & 150 $^{+ 20 } _{- 20 }$ &-0.62& - & 25 $^{+ 1 } _{- 1 }$ &-0.96& - & 120 $^{+ 10 } _{- 10 }$ &-0.62& - \\
 FCC167 & 60 $^{+ 10 } _{- 50 }$ &-1&-0.65& 36 $^{+ 3 } _{- 3 }$ &-0.8&-0.05& $\leq$ 25 &-1.3& -0.55 \\
 FCC179 & 50 $^{+ 10 } _{- 50 }$ &-1.06&-0.74& 48 $^{+ 2 } _{- 2 }$ &-0.68&-0.14& $\leq$ 5 &$\leq$-1.95& $\leq$-1.41 \\
 FCC184 & 70 $^{+ 10 } _{- 70 }$ &-0.96&-0.62& 60 $^{+ 3 } _{- 3 }$ &-0.58&0.17& $\leq$ 6 &$\leq$-1.88& $\leq$-1.14 \\
 FCC235 & 760 $^{+ 200 } _{- 210 }$ &0.09& - & $\leq$9 $^{+ 1 } _{- 1 }$ &$\leq$-1.39& - & 750 $^{+ 200 } _{- 210 }$ &0.17& - \\
 FCC263 & 210 $^{+ 40 } _{- 40 }$ &-0.48&-0.41& 90 $^{+ 10 } _{- 10 }$ &-0.42&-0.37& 120 $^{+ 20 } _{- 20 }$ &-0.63& -0.61 \\
 FCC282 & 140 $^{+ 30 } _{- 100 }$ &-0.64& - & 60 $^{+ 10 } _{- 10 }$ &-0.57& - & $\leq$ 79 &-0.81& - \\
 FCC285 & 200 $^{+ 100 } _{- 90 }$ &-0.5&-0.57& $\leq$21 $^{+ 4 } _{- 4 }$ &$\leq$-1.04&$\leq$-1.52& 170 $^{+ 90 } _{- 80 }$ &-0.46& -0.53 \\
 FCC290 & 50 $^{+ 10 } _{- 10 }$ &-1.06&-0.72& 29 $^{+ 2 } _{- 2 }$ &-0.9&-0.32& 24 $^{+ 2 } _{- 2 }$ &-1.31& -0.76 \\
 FCC306 & 8000 $^{+ 6000 } _{- 6000}$ &1.1& - & $\leq$ 4200 $^{+ 1300} _{- 1300}$ & $\leq$1.26& - & 4000 $^{+ 3000} _{- 3000}$ &0.85& - \\
 FCC308 & 100 $^{+ 20 } _{- 20 }$ &-0.79&-0.58& 37 $^{+ 3 } _{- 3 }$ &-0.79&-0.51& 62 $^{+ 9 } _{- 9 }$ &-0.91& -0.7 \\
 FCC312 & 240 $^{+ 30 } _{- 30 }$ &-0.42&-0.21& 90 $^{+ 10 } _{- 10 }$ &-0.43&-0.15& 150 $^{+ 10 } _{- 10 }$ &-0.53& -0.32 \\
 FCC335 & 450 $^{+ 120 } _{- 270 }$ &-0.14& - & 140 $^{+ 20 } _{- 20 }$ &-0.23& - & $\leq$ 311 &$\leq$-0.21& - \\
	\hline
	\end{tabular}
	\textit{Notes:} \textbf{1}: Object name according to the Fornax Cluster Catalogue; \textbf{2}: Total gas-to-dust ratio ($\left(\text{M}_{\text{H}\textsc{i}}+\text{M}_{\text{H}_2}\right)/\text{D}$); \textbf{3}: Total gas-to-dust residual compared to the DustPedia median as a function of stellar mass (see Figure \ref{fig:total_GD_SM}); \textbf{4}: Total gas-to-dust residual compared to the DustPedia median as a function of metallicity (see Figure \ref{fig:GD_metal_total}); \textbf{5}: H$_2$-to-dust ratio; \textbf{6}: H$_2$-to-dust residual compared to the DustPedia median as a function of stellar mass (see Figure \ref{fig:GD_SM}); \textbf{7}: H$_2$-to-dust residual compared to the DustPedia median as a function of metallicity (see Figure \ref{fig:GD_metal}); \textbf{8}: \HI -to-dust ratio; \textbf{9}: \HI -to-dust residual compared to the DustPedia median as a function of stellar mass (see Figure \ref{fig:GD_SM_HI}); \textbf{10}: \HI -to-dust residual compared to the DustPedia median as a function of metallicity (see Figure \ref{fig:GD_metal_HI}).
	\end{threeparttable}
\end{table*}

\firstround{In order to compare the dust and gas in our cluster galaxies to those in the field, we use data from the DustPedia project.} DustPedia\footnote{\url{http://dustpedia.astro.noa.gr/}} \citep{Davies2017} covers all 875 extended galaxies within 3000 \kms\ observed by the \textit{Herschel} Space Observatory. Here, we use a sub-sample of DustPedia, consisting of the 209 galaxies for which \MH2\ measurements are available. This sample includes the \textit{Herschel} Virgo Cluster Survey (HeViCS, \citealt{Davies2010}) sample from \citet{Corbelli2012}. \changes{After eliminating Fornax galaxies,} we split up this sample in Virgo galaxies and field galaxies. Galaxies are defined to be in the Virgo cluster if they are within twice the virial radius of the Virgo cluster, assumed to be 1.7 Mpc \citep{Fukushige2001}. \firstround{This corresponds to $\sim 5.7^\text{o}$ at the distance to the Virgo cluster, here assumed to be 16.5 Mpc \citep{Mei2007}. Distances to individual DustPedia galaxies, \secround{adopted from the DustPedia database,} are redshift-independent distance estimates from HyperLEDA where available, and redshift-independent estimates from the NASA/IPAC Extragalactic Database (NED) if not \secround{(more details on redshift-independent distance estimates by HyperLEDA and NED can be found in \citealt{Makarov2014} and \citealt{Steer2017}, respectively)}. If both are unavailable, \secround{bulk} flow-corrected redshift-derived values provided by NED are used, assuming a Hubble constant of $H_0 = 73.24$ \kms\ Mpc$^{-1}$ \citep{Riess2016}. Finally, we create an additional sub-sample of Virgo galaxies located inside the cluster virial radius.}

H$_2$ masses for DustPedia galaxies were compiled and homogenised from a wide variety of sources by \citet{Casasola2020}. We use their \MH2\ estimates that were derived using a fixed \xco, which we recalibrate to match the metallicity-dependent prescription used in this work.

We adopt dust masses estimated using a modified blackbody model, scaled to match the emissivity of $\kappa = 0.192 \frac{\text{m}^2}{\text{kg}}$ at 350 $\mu$m, used to estimate dust masses of the Fornax sample, and a $\beta$-value of 1.790 \citep{Nersesian2019}. 

\refrep{To maximise consistency with our Fornax sample, rather than adopting published metallicities, we apply the DOP16 calibration to the (extinction corrected) \textit{line ratios} from \citet{Vis2019}}. We average metallicities from all detected star forming regions for each galaxy, consistent with the spatially averaged metallicities used to estimate global metallicities in the Fornax sample. Details on the line flux measurements of the DustPedia sample can be found in \citet{Vis2019} and on the DustPedia website.

DustPedia \HI\ fluxes were compiled from the literature (Casasola et al., in prep.) and converted to $\text{M}_{\text{H}{\textsc i}}$ using 
\begin{equation}
\label{eq:HI_mass}
\text{M}_{\text{H}{\textsc i}} = 2.36 \times 10^5 f_{\text{H}{\textsc i}}\ D^2,
\end{equation}
where $f_{\text{H}{\textsc i}}$ is the compiled \HI\ flux in Jy km s$^{-1}$ and D the best distance measurement from \citet{Clark2018} in Mpc. More details and references can be found in \citet{Vis2019}.


\refrep{Stellar masses of the DustPedia sample were taken from z0MGS, as described in \S \ref{sub:stellar_masses}. This means that the stellar masses of the DustPedia sample are fully consistent with the vast majority of the Fornax sample, with the exception of only the two objects discussed in \S \ref{sub:stellar_masses}.}

Thus, as explained above, the dust, H$_2$, \HI, and stellar masses we use here are calculated identically for both our Fornax and the DustPedia comparison sample, and so can be directly compared. 


\subsection{Other literature samples}
\label{sub:other_lit_samples}
\firstround{There are several other literature samples which have molecular gas, atomic gas, and dust masses available, as well as stellar masses and/or metallicities. These include the samples compiled by \citet{RemyRuyer2014}, such as the Dwarf Galaxy Survey (DGS, \citealt{Madden2013}), and Key Insights on Nearby Galaxies: a Far-Infrared Survey with \textit{Herschel} (KINGFISH, \citealt{Kennicutt2011}), the sub-sample of LITTLE THINGS from Cigan et al., in prep., and several others. Unfortunately, it has proven impossible to recalibrate these samples to rely on the same assumptions as the Fornax and DustPedia samples studied here. Therefore, they were not included in this work. Homogenising these datasets to follow the same assumptions used here (as far as is possible) suggests our conclusions above also hold in comparison to these samples.}





\section{Analysis}
\label{sec:results}

\subsection{Gas-to-dust ratios}
\label{sub:GD_SM}
\begin{figure*}
	\centering
	\includegraphics[width=0.66\textwidth]{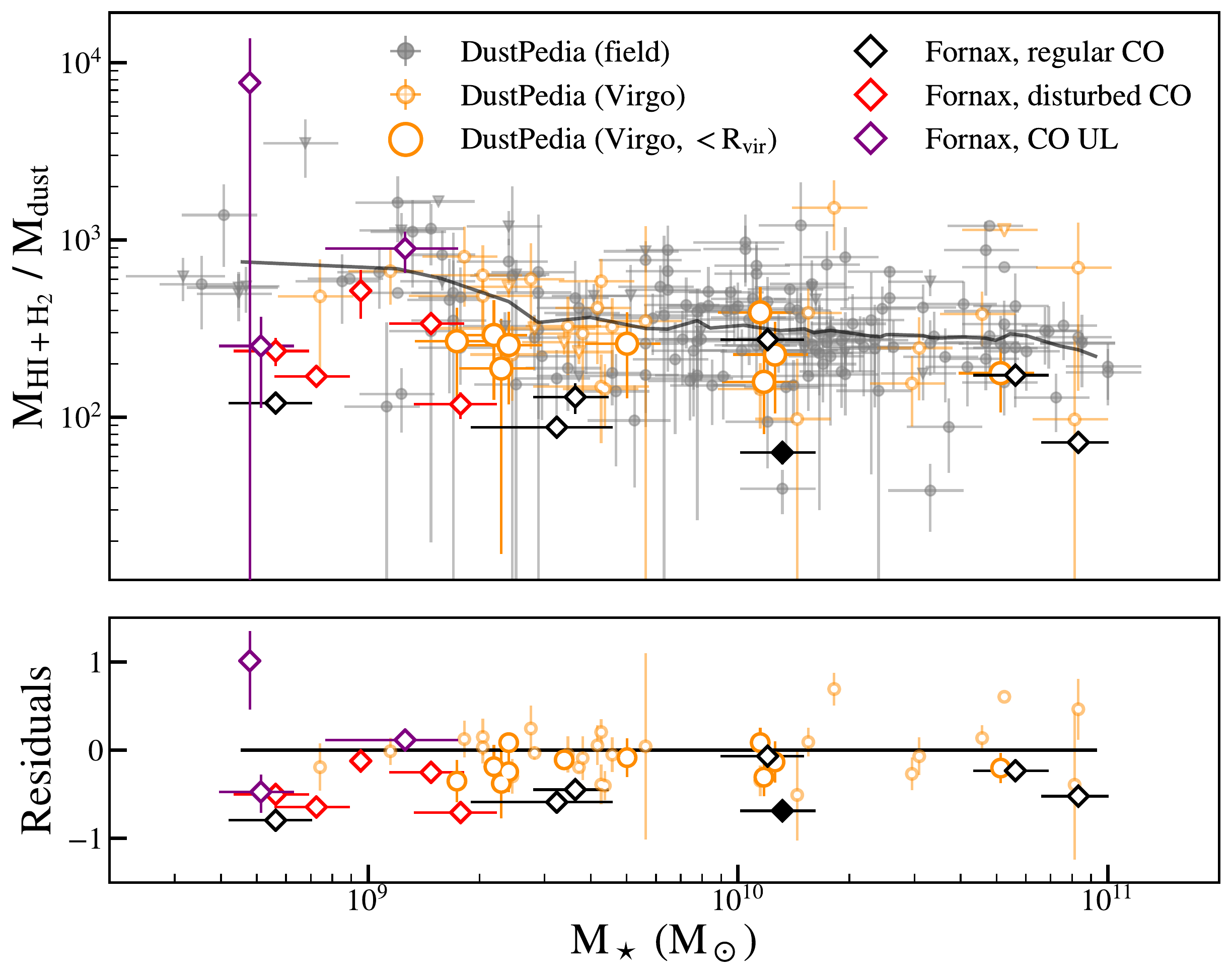}
	\caption{\textit{Upper panel}: Total gas-to-dust (\HI\ + H$_2$) ratios in the Fornax cluster compared to those in the DustPedia \citep{Davies2017} field (grey) and Virgo (orange) samples. Virgo galaxies inside $R_\text{vir}$ are highlighted with larger markers. Fornax galaxies are indicated with diamond-shaped markers, in black for galaxies with regular CO emission and in red for galaxies with disturbed CO emission as classified in Z19 (upper limits are shown in purple). \refrep{NGC1436 (see \S \ref{sub:NGC1436}) is indicated with a filled diamond-shaped marker.} Upper limits are indicated with downward-facing triangles for all samples. The solid grey line indicates the rolling median of the DustPedia field sample 
\textit{Lower panel}: Residuals of gas-to-dust ratios in Fornax and Virgo cluster galaxies compared to the rolling median of DustPedia field galaxies, shown as a solid line, or a dashed line where extrapolated. \firstround{Note that, for the purpose of visibility, the bottom panel is set to show values between -1.5 and 1.5, which causes the strongly discrepant galaxy FCC067 (NGC1351A) to fall off the plot in some of the following figures. According to KS tests, Fornax galaxies and Virgo galaxies inside $R_\text{vir}$ have systematically lower gas-to-dust ratios than field galaxies at fixed stellar mass (see Table \ref{tab:KS_results}).}}
	\label{fig:total_GD_SM}
\end{figure*}

\begin{figure*}
	\centering	
	\includegraphics[width=0.66\textwidth]{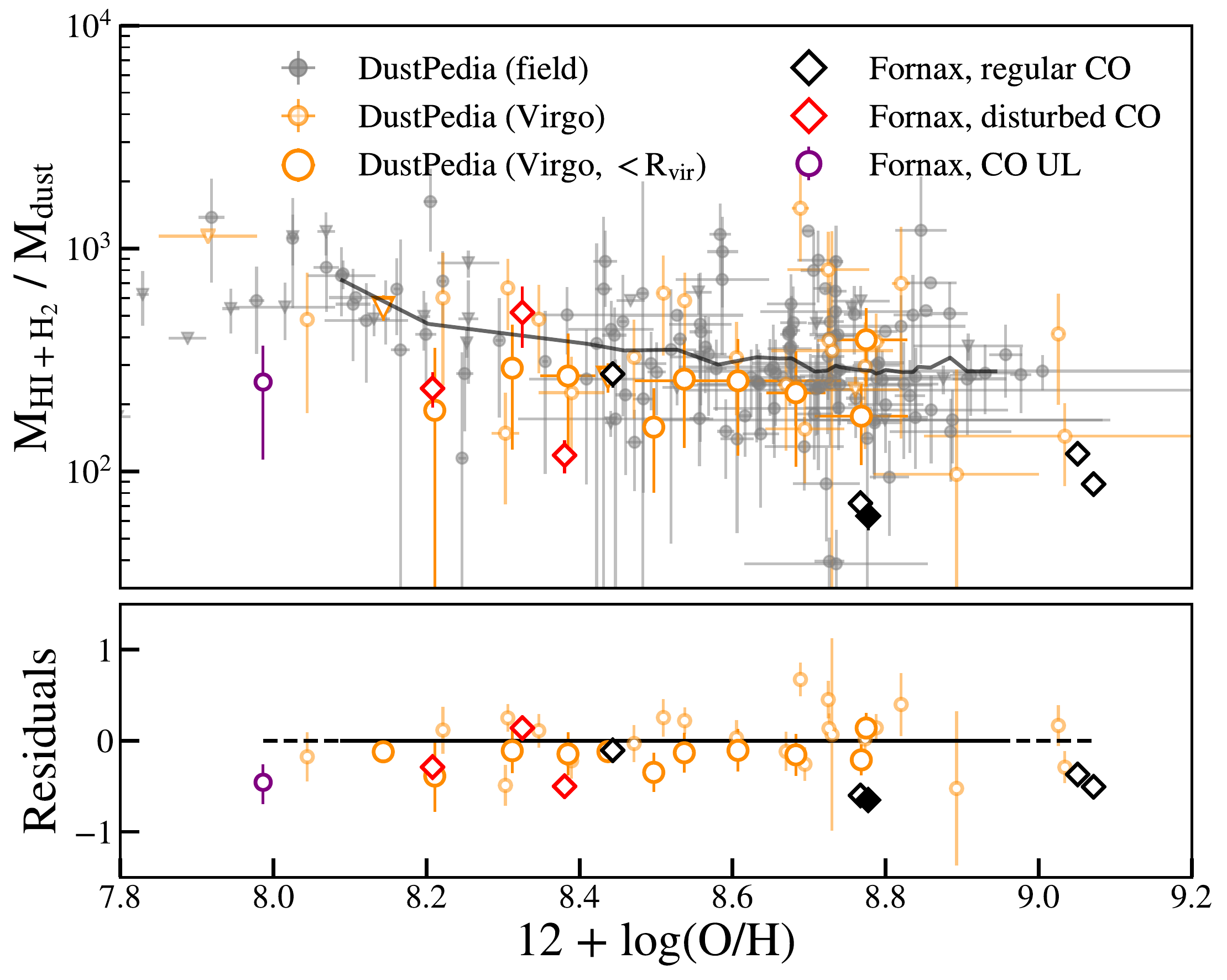}
	\caption{Similar to Figure \ref{fig:total_GD_SM}, showing total gas-to-dust ratios as a function of metallicity \secround{for a sub-sample of Fornax galaxies for which MUSE data is available from F3D (see Table \ref{tab:targets_fornax})}. Fornax cluster galaxies and Virgo galaxies inside $R_\text{vir}$ have systematically lower gas-to-dust ratios than field samples.}
	\label{fig:GD_metal_total}
\end{figure*}

\begin{figure*}
	\centering
	\includegraphics[width=0.65\textwidth]{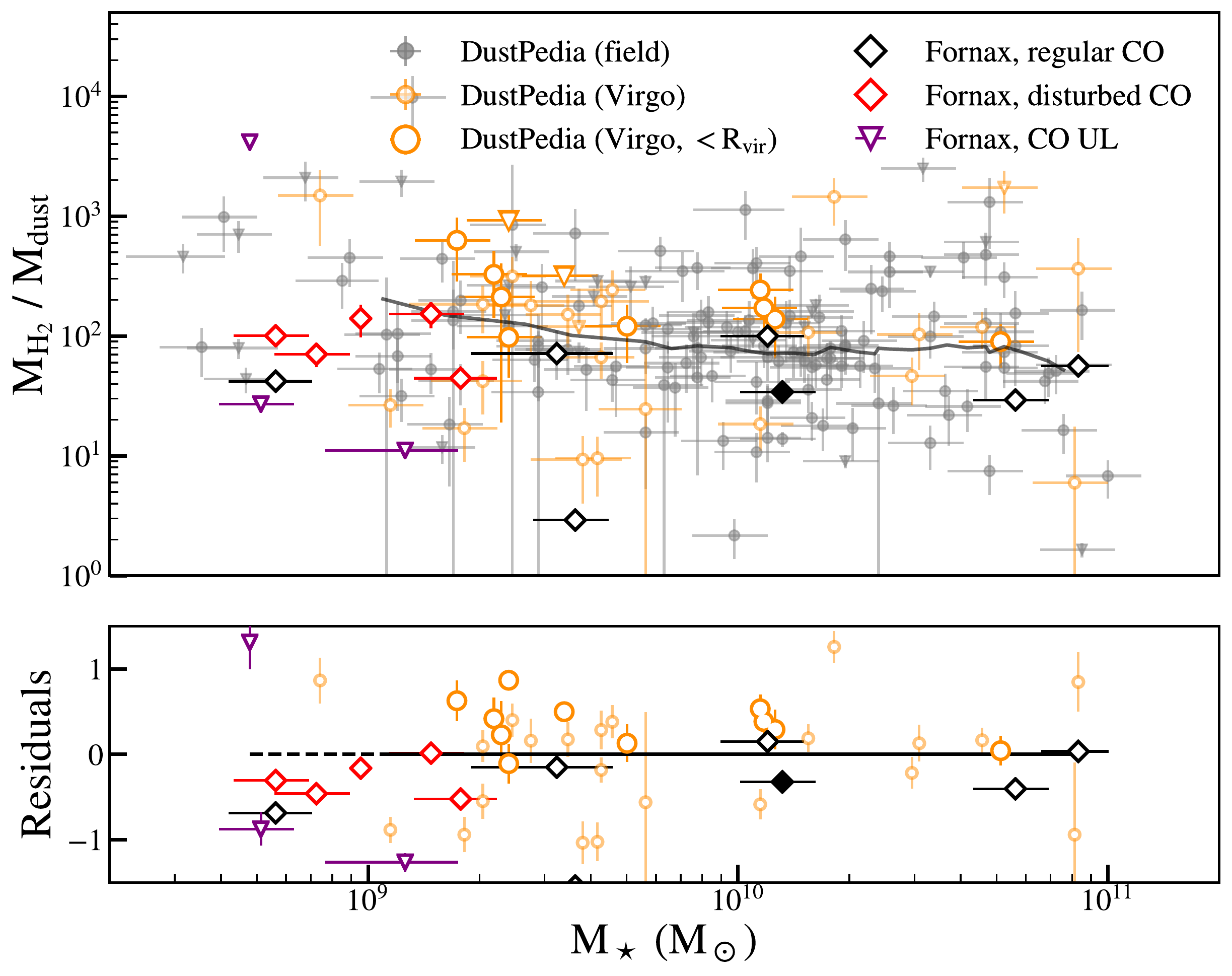}
	\caption{Similar to Figure \ref{fig:total_GD_SM}, but showing molecular gas-to-dust ratios plotted as a function of stellar mass. Fornax cluster galaxies have systematically lower molecular gas-to-dust ratios than field galaxies, whereas Virgo galaxies inside $R_\text{vir}$ show \textit{in}creased H$_2$-to-dust ratios.}
	\label{fig:GD_SM}
\end{figure*}

\begin{figure*}
	\centering
	\includegraphics[width=0.67\textwidth]{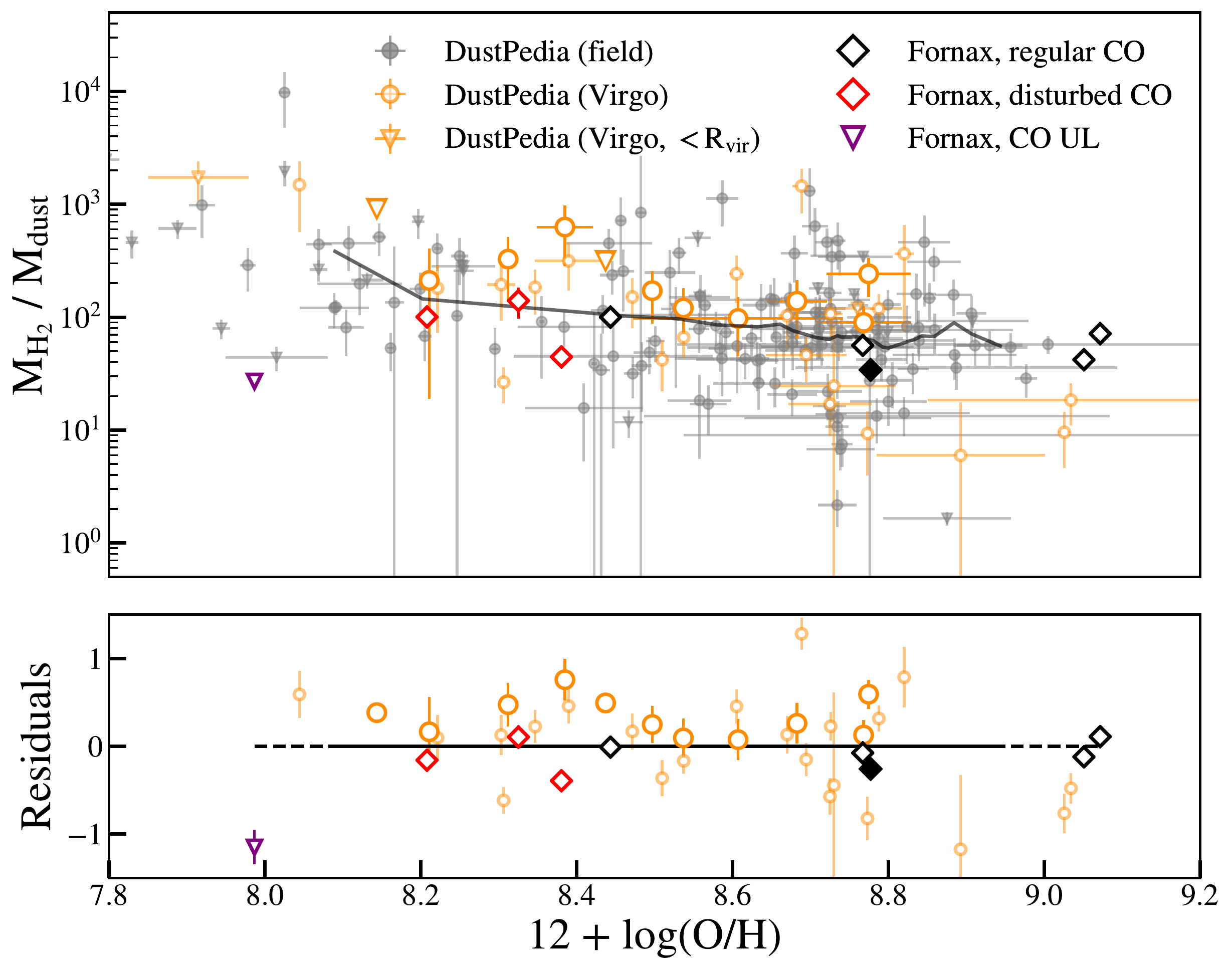}		
	\caption{Similar to Figure \ref{fig:GD_metal_total}, but showing molecular gas-to-dust ratios as a function of metallicity. Molecular gas-to-dust ratios of Fornax galaxies are systematically lower than those of field galaxies at fixed metallicity, while those inside the virial radius of the Virgo cluster are \textit{in}creased.}
	\label{fig:GD_metal}
\end{figure*}

\begin{figure*}
	\centering
	\includegraphics[width=0.67\textwidth]{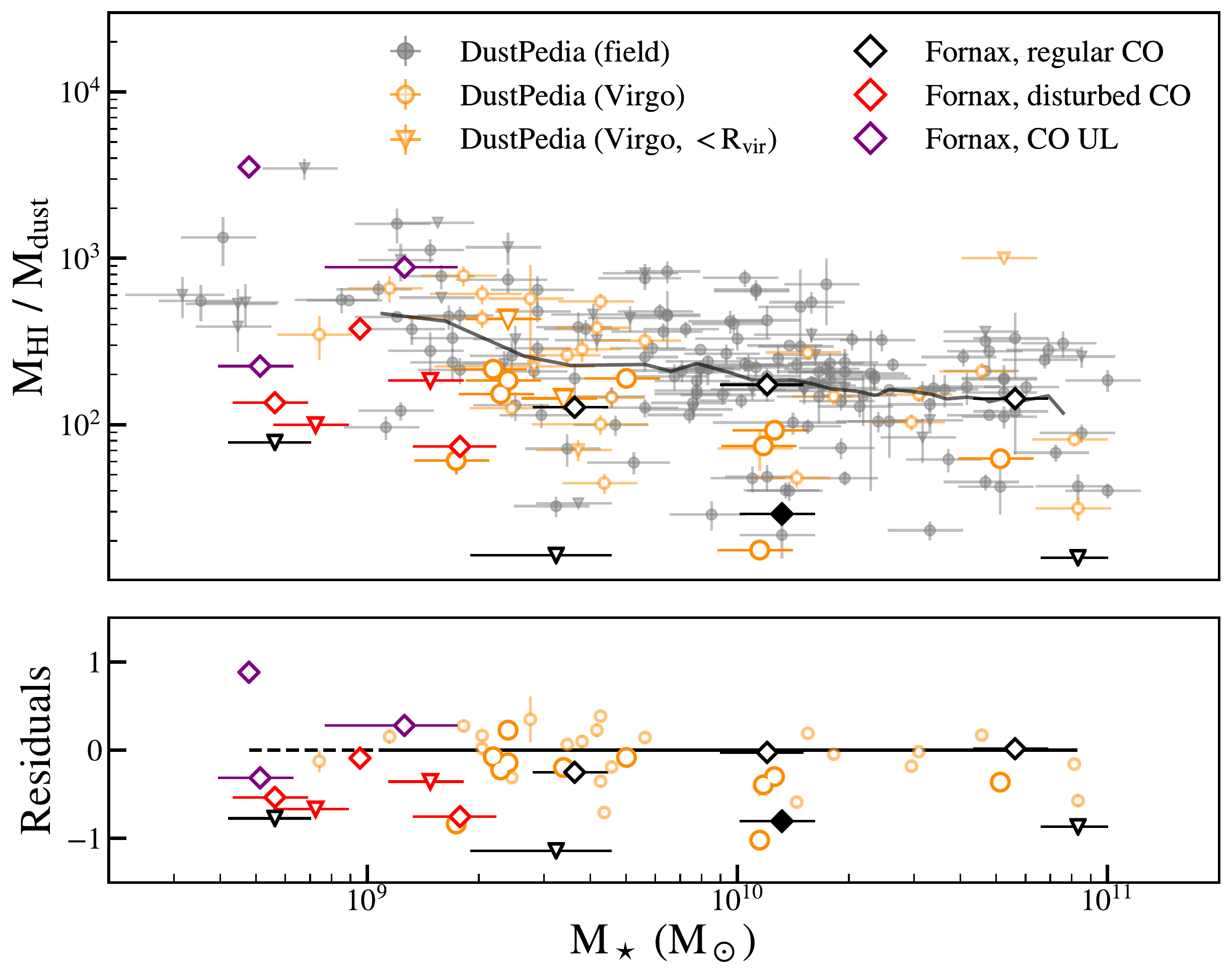}
	\caption{Similar to Figure \ref{fig:total_GD_SM}, but showing \HI -to-dust ratios as a function of stellar mass. Fornax galaxies and Virgo galaxies inside the virial radius have systematically lower \HI -to-dust ratios than galaxies in the field at fixed stellar mass. A possible decrease is also seen in Virgo galaxies between 1 and 2 $R_\text{vir}$.}
	\label{fig:GD_SM_HI}
\end{figure*}

\begin{figure*}
	\centering
	\includegraphics[width=0.67\textwidth]{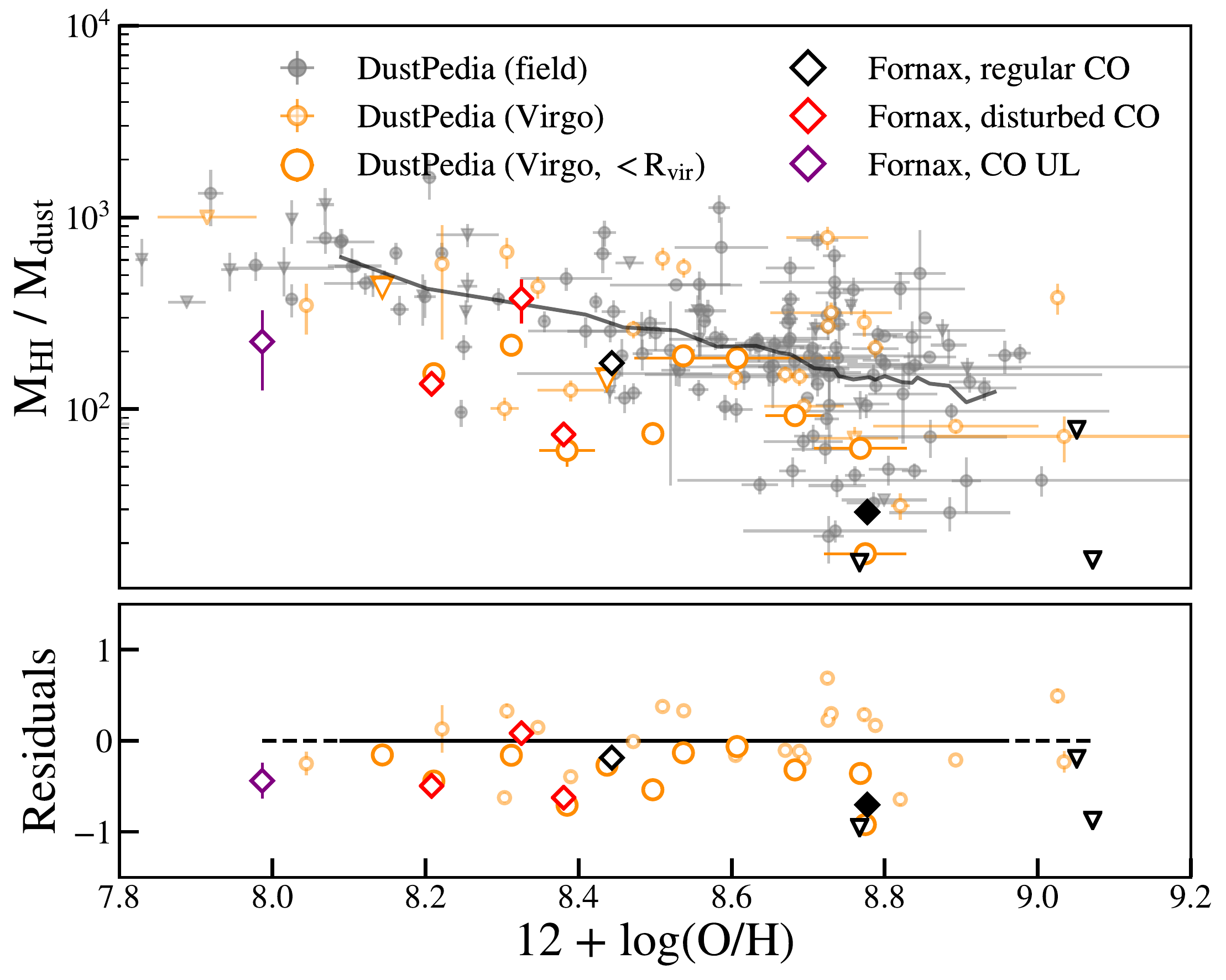}		
	\caption{Similar to Figure \ref{fig:GD_metal_total}, but showing \HI -to-dust ratios as a function of metallicity. These are systematically lower in Fornax cluster galaxies and Virgo galaxies inside $R_\text{vir}$ compared to field galaxies at fixed metallicity.}
	\label{fig:GD_metal_HI}
\end{figure*}

\firstround{In order to examine the effect of the Fornax cluster environment on the gas and dust in galaxies, we construct gas-to-dust ratios for our sample galaxies (listed in Table \ref{tab:gas-to-dust_ratios}). We also tabulate the offset between the gas-to-dust ratio of each Fornax galaxy and the median gas-to-dust ratio of field galaxies at the same stellar mass, and metallicity \secround{for the nine galaxies for which MUSE data from F3D is available,} (calculated using the DustPedia field galaxy sample). For comparison, we also include the DustPedia Virgo galaxies in our analysis. Total gas-to-dust ratios are shown in Figures \ref{fig:total_GD_SM} and \ref{fig:GD_metal_total}, as a function of stellar mass, and metallicity, respectively. Similarly, Figures \ref{fig:GD_SM} and \ref{fig:GD_metal} show the H$_2$-to-dust ratio, and Figures \ref{fig:GD_SM_HI} and \ref{fig:GD_metal_HI} the \HI -to-dust ratio. Fornax galaxies are plotted as diamond-shaped markers. Galaxies with regular molecular gas reservoirs are shown in black, and those with disturbed molecular gas reservoirs in red (as classified in Z19). CO upper limits (for which the molecular gas morphology is unknown) are shown in purple. \firstround{Virgo galaxies are shown in orange, with galaxies inside the virial radius highlighted with larger markers}. The DustPedia field sample is shown with grey markers. Upper limits are shown as downward-facing triangles for all samples. The solid, grey line represents the rolling median of the DustPedia field sample, which is calculated using bins with a fixed number of 10 galaxies, overlapping by half a bin size. The bottom panels in these figures show the residuals of Fornax and Virgo galaxies compared to this median, which is shown as a black solid line, or a black dashed line where extrapolated.}



\refrep{In Figures \ref{fig:total_GD_SM} and \ref{fig:GD_metal_total} we can see that total gas-to-dust ratios in the Fornax cluster are low compared to the field, both at fixed stellar mass, and metallicity. This also applies to galaxies in the Virgo cluster, especially inside its virial radius. Figures \ref{fig:GD_SM} and \ref{fig:GD_metal} show that molecular gas-to-dust ratios are also decreased in the Fornax cluster compared to the field. However, galaxies in the Virgo cluster show the opposite result: they are \secround{(marginally, but systematically)} increased compared to the field. Finally, Figures \ref{fig:GD_SM_HI} and \ref{fig:GD_metal_HI} show that \HI -to-dust ratios are decreased in both the Fornax and Virgo clusters, although the difference between the Virgo cluster and the field is smaller if galaxies located inside $2 R_\text{vir}$ are included. In the Fornax cluster there is no significant difference in gas-to-dust ratio between galaxies with regular and disturbed CO reservoirs in any of the Figures.}

\refrep{We quantify the results described above by applying Kolmogorov-Smirnov (KS) tests and Anderson-Darling (AD) tests to the residuals of the Fornax and Virgo data compared to DustPedia in all six Figures. Because these tests do not take into account uncertainties, we use a Monte Carlo approach to make sure these are implemented in the final result. We perturb each value by a random number drawn from a normal distribution with $\mu$ the measured value and $\sigma$ the associated uncertainty. This is done $10^6$ times, after which the $\mu$ and $\sigma$ of the resulting distribution of the KS and AD statistics are adopted as the test results. These are are summarised in Table \ref{tab:KS_results}. The KS statistic is between 0 and 1, and the closer to 1 it is, the less likely it is that both samples are drawn from the same distribution. The AD test returns a statistic (referred to as the A2 statistic) as well as corresponding critical values at a discrete number of confidence intervals. The null-hypothesis that both samples are drawn from the same distribution can be rejected with a certain probability if the difference between the A2 statistic and the critical level is positive at the corresponding confidence level. In Table \ref{tab:KS_results} we report the difference between the A2 statistic and the critical value at 0.1\%.}

\refrep{Both tests return similar results in all cases. In almost all cases, the null-hypothesis that the Fornax and Virgo samples are drawn from the same distribution as the DustPedia field sample can be rejected. The exception is the Fornax \HI\ mass as a function of stellar mass, for which the spread in both statistics is too large to draw strong conclusions.}

\begin{table*}
	\centering
	\begin{threeparttable}
	\caption{Results of a Monte Carlo analysis of KS and AD tests applied to the residuals (data of the respective clusters compared to the rolling median of the DustPedia field sample) in Figures \ref{fig:total_GD_SM} through \ref{fig:GD_metal_HI}.}
	\label{tab:KS_results}
	\setlength{\tabcolsep}{1.5mm}
	\begin{tabular}{rrrrrrrrrrrrrr}
	\hline
	& & & & \multicolumn{2}{c}{Fornax} & & \multicolumn{2}{c}{Virgo (inside $R_\text{vir}$)} & & \multicolumn{2}{c}{Virgo (inside 2$R_\text{vir}$)} \\
	\cline{5 - 6} \cline{8 - 9} \cline{11 - 12}
	\rule{0pt}{3ex}  
	Figure & x-axis & y-axis & & D statistic & A2 - crit. & & D statistic & A2 - crit. & & D statistic & A2 - crit. \\
	(1) & (2) & (3) & & (4) & (5) & & (6) & (7) & & (8) & (9) \\
	\cline{1 - 3} \cline{5 - 6} \cline{8 - 9} \cline{11 - 12}
	\ref{fig:total_GD_SM} & M$_\star$ & M$_{\text{H\textsc{i}} + \text{H}_2}$ & & 0.98 $\pm$ 0.005 & 28 $\pm$ 1 & & 0.98 $\pm$ 0.05 & 22 $\pm$ 2 & & 0.98 $\pm$ 0.06 & 59 $\pm$ 6 \\
	\ref{fig:GD_SM} & M$_\star$ & M$_{\text{H}_2}$ & & 0.89 $\pm$ 0.01 & 17 $\pm$ 2 & & 0.89 $\pm$ 0.07 & 8 $\pm$ 3 & & 0.87 $\pm$ 0.12 & 31 $\pm$ 9 \\
	\ref{fig:GD_SM_HI} & M$_\star$ & M$_\text{H\textsc{i}}$ & & 0.79 $\pm$ 0.28 & 12 $\pm$ 12 & & 0.99 $\pm$ 0.14 & 20 $\pm$ 6 & & 0.99 $\pm$ 0.15 & 49 $\pm$ 12 \\
	\ref{fig:GD_metal_total} & 12 + log(O/H) & M$_{\text{H\textsc{i}} + \text{H}_2}$ & & 0.99 $\pm$ 0.03 & 17 $\pm$ 1 & & 0.99 $\pm$ 0.04 & 20 $\pm$ 2 & & 0.99 $\pm$ 0.04 & 46 $\pm$ 4 \\
	\ref{fig:GD_metal} & 12 + log(O/H) & M$_{\text{H}_2}$ & & 0.99 $\pm$ 0.006 & 17 $\pm$ 1 & & 0.99 $\pm$ 0.04 & 20 $\pm$ 2 & & 0.98 $\pm$ 0.06 & 46 $\pm$ 4 \\
	\ref{fig:GD_metal_HI} & 12 + log(O/H) & M$_\text{H\textsc{i}}$ & & 0.99 $\pm$ 0.04 & 17 $\pm$ 2 & & 0.99 $\pm$ 0.09 & 20 $\pm$ 4 & & 0.99 $\pm$ 0.08 & 46 $\pm$ 6 \\
	\hline
	\end{tabular}
	\textit{Notes:} \textbf{1}: reference to the Figure showing the data the KS test is applied to; \textbf{2}: x-axis of that Figure; \textbf{3}: y-axis of that Figure; \textbf{4, 6, 8}: D-statistic resulting from the KS test applied to the Fornax, Virgo inside $R_\text{vir}$, and Virgo inside $2 R_\text{vir}$ data, respectively, including uncertainties from a Monte Carlo analysis; \textbf{5, 7, 9}: difference between the A2 statistic and critical value at the 0.1\% confidence level resulting from an Anderson-Darling test applied to the Fornax, Virgo inside $R_\text{vir}$, and Virgo inside $2 R_\text{vir}$ data respectively;
	\end{threeparttable}
\end{table*}

\subsection{Variation with cluster-centric distance}
\label{subsec:cluster_centric_distance}
\begin{figure*}
	\centering
	\includegraphics[width=0.7\textwidth]{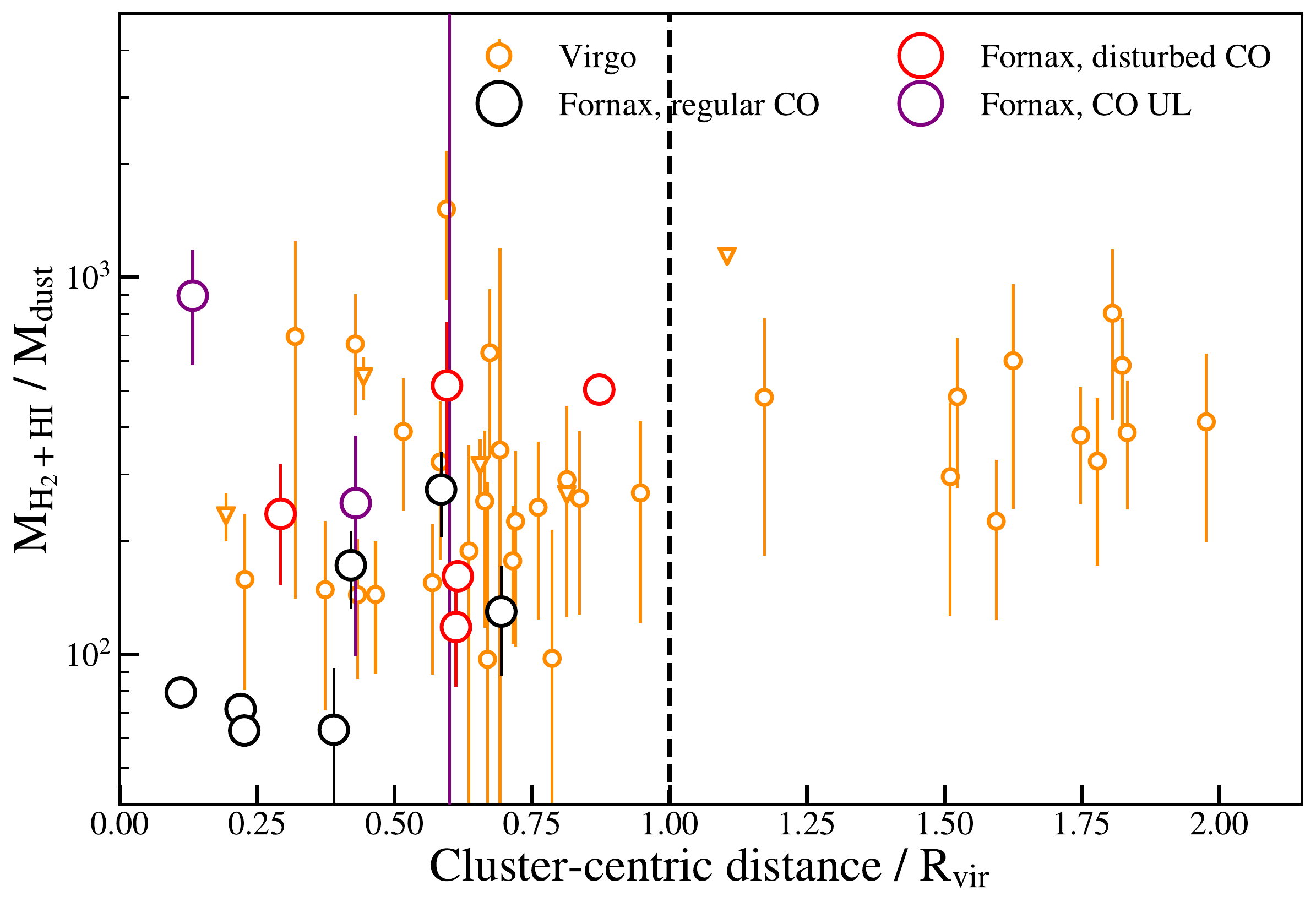}
	\caption{\changes{Total gas-to-dust ratios as a function of relative (projected) distance to the cluster centre are shown with colours and markers similar to those in Figure \ref{fig:GD_SM}. \firstround{The virial radius is indicated with a black dashed line.} No significant gradient in gas-to-dust fractions is observed in any of the clusters, although there is possibly a slight increase in gas-to-dust radio with cluster-centric radius in the Virgo cluster (a Kendall's Tau test returns a tau statistic of 0.23 and a p-value of 0.04).}}
	\label{fig:GD_distance}
\end{figure*}

In Figure \ref{fig:GD_distance} we show gas-to-dust ratios as a function of projected cluster-centric distance for both clusters, which are represented by the same colours as in previous Figures. These distances are normalised by each cluster's virial radius, \firstround{indicated by a black dashed line}. There is no clear relation between gas-to-dust ratio and cluster-centric radius for either of the clusters, although there is a possible increase in gas-to-dust ratio with cluster-centric radius in the Virgo cluster (a Kendall's Tau test returns a tau statistic of 0.23 and a p-value of 0.04). In the Virgo cluster, there is evidence that \HI\ deficiencies decrease with cluster-centric radius (e.g. \citealt{Haynes1986}, \citealt{Gavazzi2005}), which could explain this possible trend. If \HI\ deficiencies were the main driver of the decreased gas-to-dust ratios in the Fornax cluster, we might expect to also see such a trend in this cluster. However, no such trend is seen in Figure \ref{fig:GD_distance}, \firstround{although we only have a small sample of Fornax cluster objects}.

\subsection{Gas and dust fractions}
\label{sub:fractions}
\begin{figure*}

	\centering
	
	\subfloat[\label{subfig:dust_SM}]
	{\includegraphics[width=0.65\textwidth]{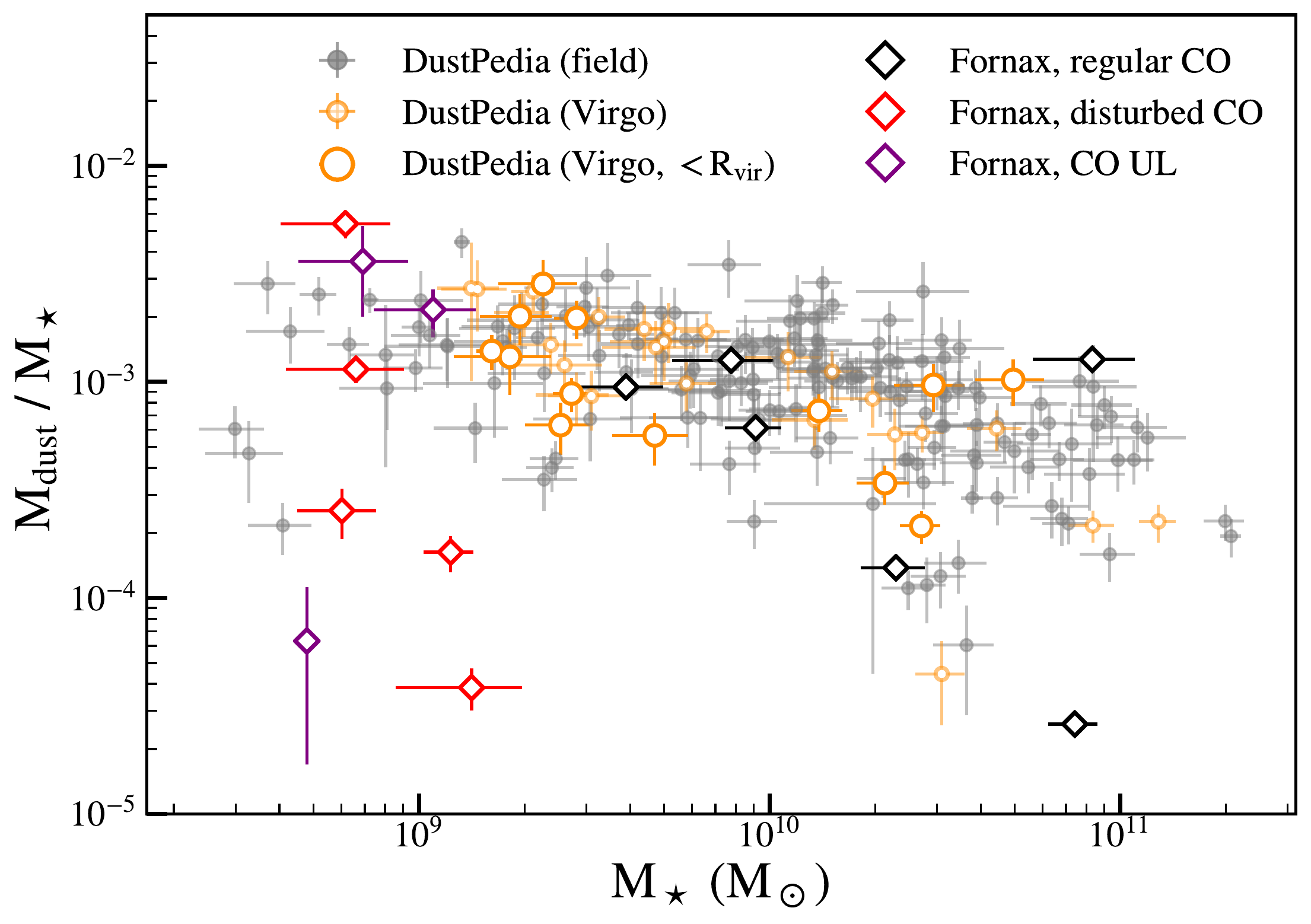}}	
	
	\subfloat[\label{subfig:gas_SM}]
	{\includegraphics[width=0.65\textwidth]{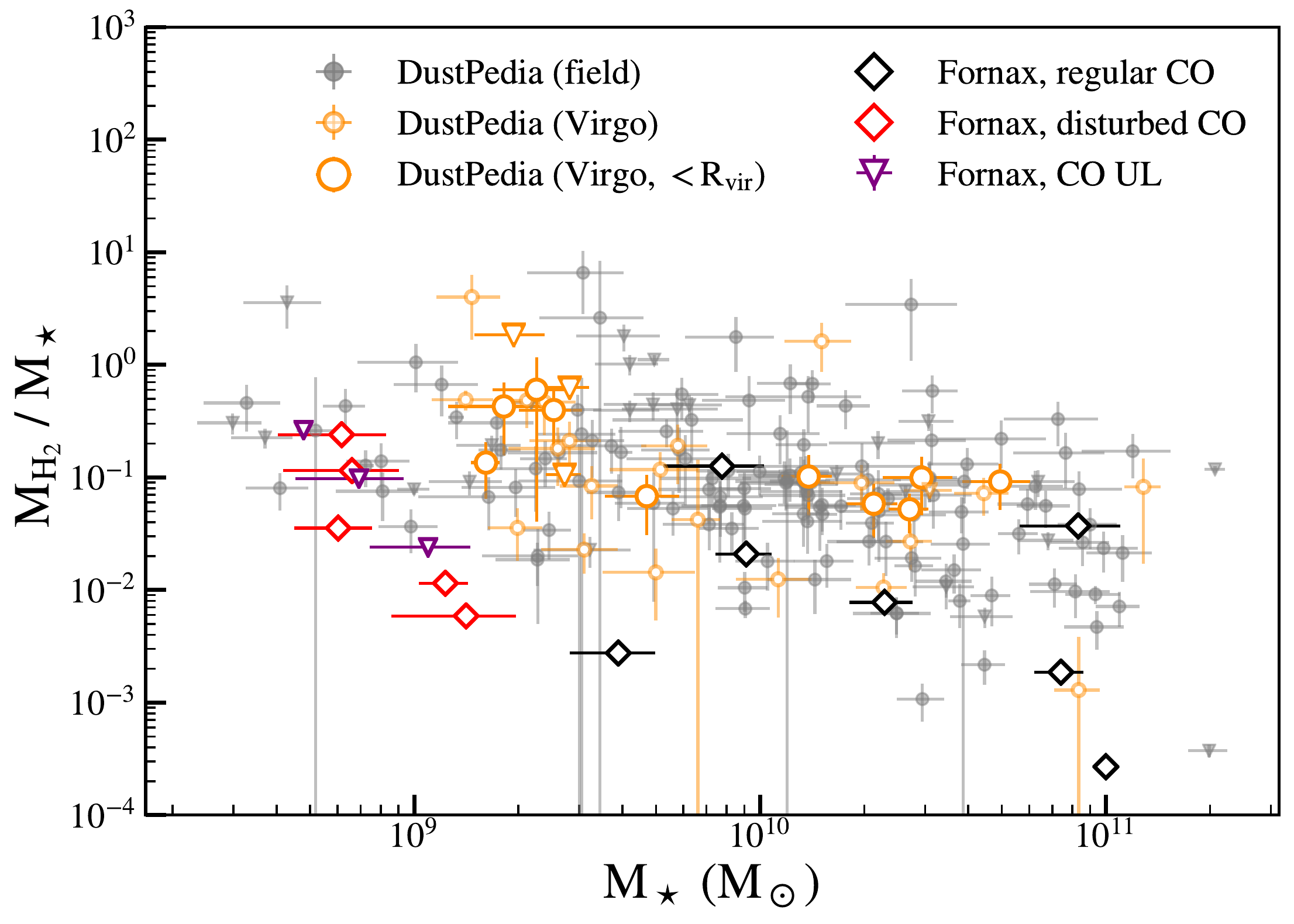}}
	
	\caption{Dust- and molecular gas-to-stellar mass fractions (panel a and panel b, respectively) in the Fornax cluster compared to the field and the Virgo cluster. Samples and symbols are the same as in previous figures. While both are low, molecular gas-to-stellar mass fractions in the Fornax cluster are especially low compared to the field and the Virgo cluster. \firstround{The dust content in the Virgo cluster is normal to slightly decreased, whereas molecular gas-to-stellar mass fractions are normal to slightly increased, in particular at low stellar mass.}}
	\label{fig:dust_gas_stellar_mass}
	
\end{figure*}

To better understand whether the decreased H$_2$-to-dust ratios in the Fornax cluster are mostly the result of increased dust fractions, or decreased H$_2$ fractions, we show these as a function of stellar mass in Figure \ref{fig:dust_gas_stellar_mass} (Figure \ref{subfig:dust_SM} shows dust fractions and Figure \ref{subfig:gas_SM} H$_2$ fractions). Markers and symbols are the same as in previous figures. The dust content in most Fornax galaxies is normal to low \secround{compared to the DustPedia field sample}, with the exception of a few dwarf galaxies, and NGC1365 at the high-mass end. The scatter in the H$_2$ fractions in Figure \ref{subfig:gas_SM} is similar to that in Figure \ref{subfig:dust_SM}, however there is a more pronounced systematic offset towards lower H$_2$ fractions in the Fornax cluster compared to the field. \firstround{This suggests the difference seen in the Fornax cluster is driven by the lower H$_2$ content of these galaxies. In Virgo dust fractions are normal to slightly decreased inside the virial radius, whereas H$_2$ fractions are increased, especially in low-stellar mass galaxies. Outside $R_\text{vir}$ no difference with the DustPedia field sample is seen.}

\subsection{Dust-to-metal ratios}
\label{sub:metal-to-dust}
\begin{figure*}
	\centering
	\subfloat[\label{subfig:metalmass_SM}]
	{\includegraphics[width=0.65\textwidth]{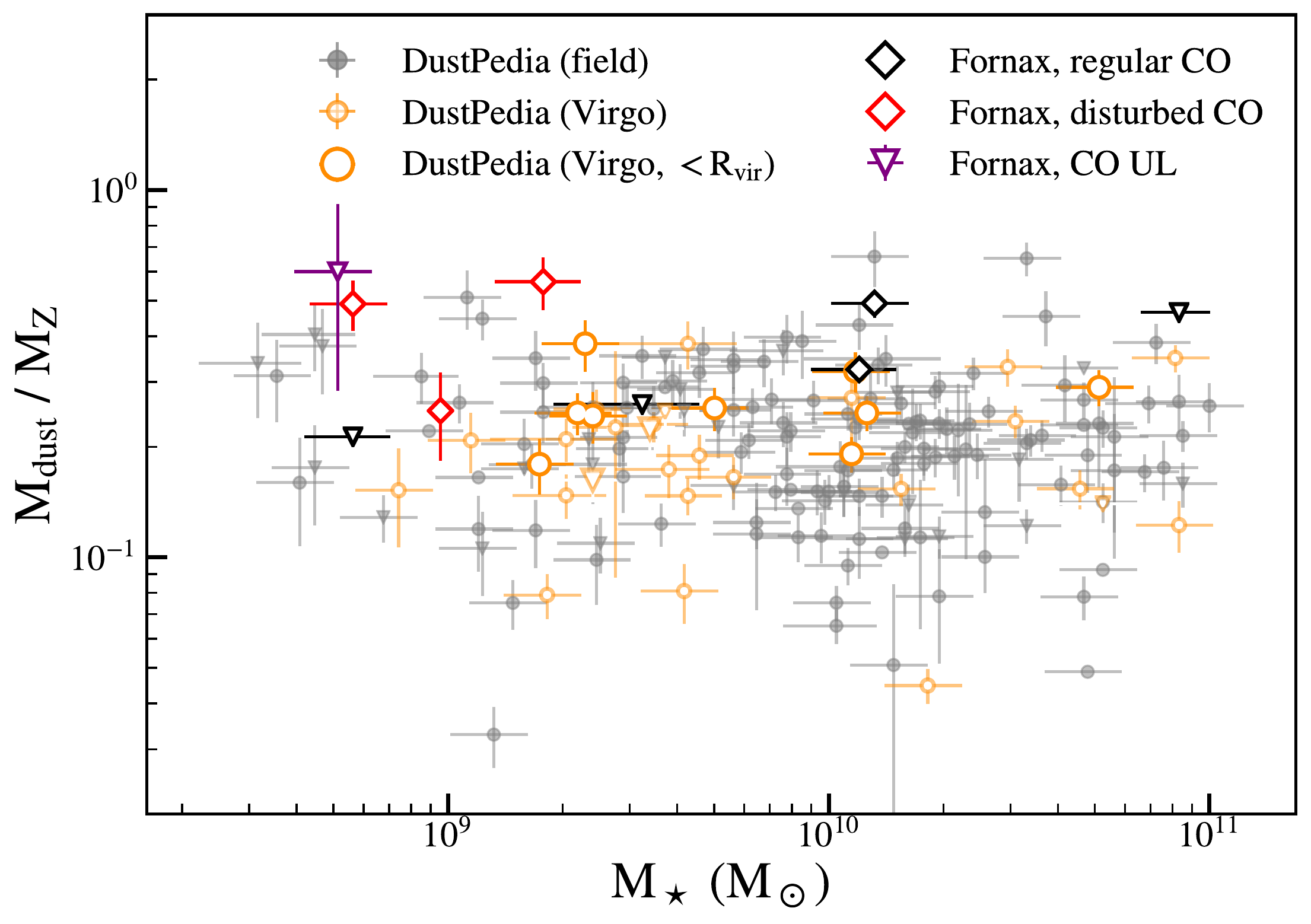}}	
	
	\subfloat[\label{subfig:metalmass_metal}]
	{\includegraphics[width=0.65\textwidth]{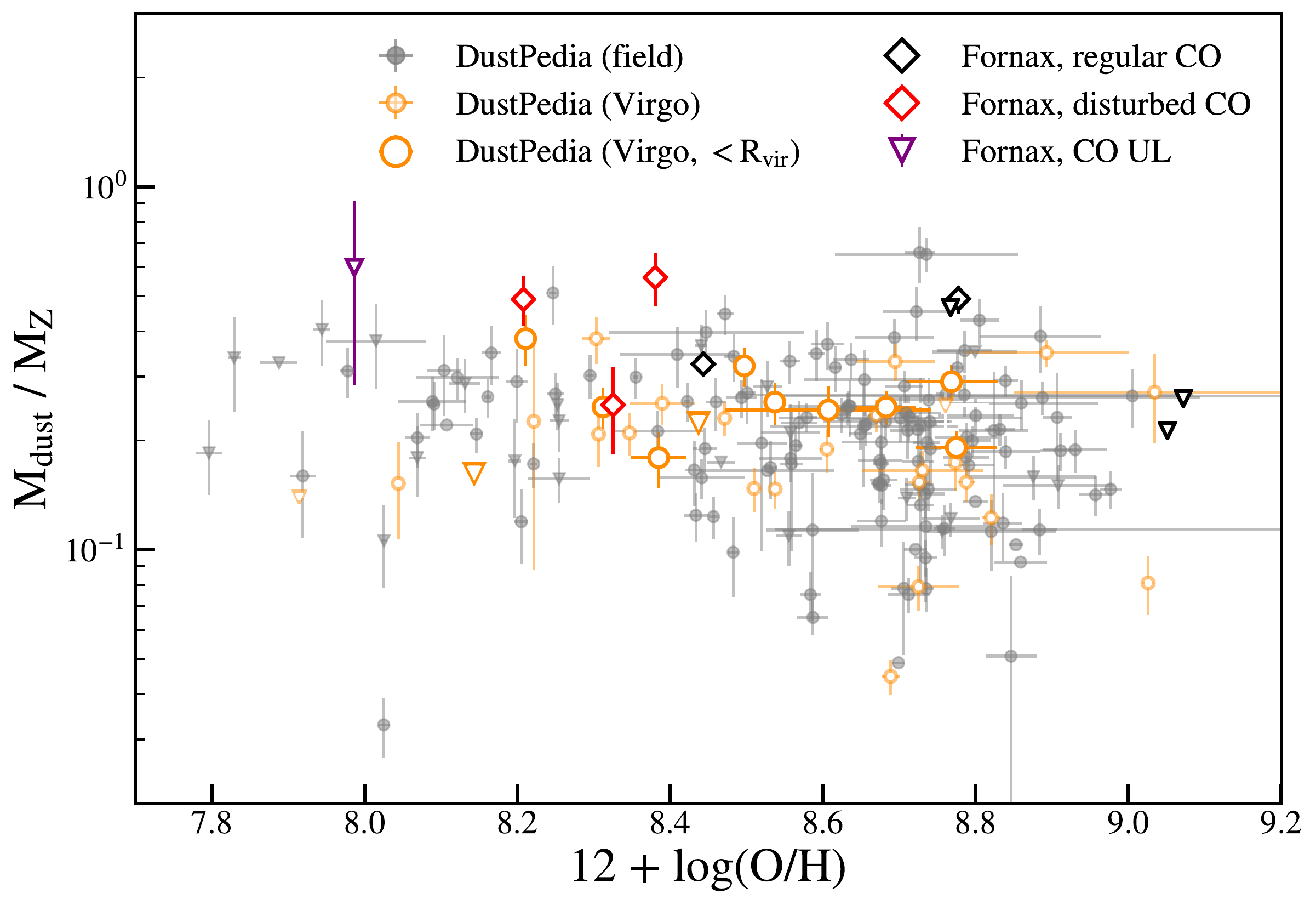}}
	
	\caption{Dust-to-metal ratios as a function of stellar mass (panel a) and metallicity (panel b). Samples and symbols are the same as in previous figures. Dust-to-metal ratios in Fornax cluster galaxies are significantly higher than those in field galaxies of similar mass and metallicity, which means that a relatively large fraction of the metals in these galaxies are locked up in dust grains. Virgo galaxies inside $R_\text{vir}$ also show a slight, but less significant, increase in dust-to-metal ratio at fixed stellar mass. This difference is less clear at fixed metallicity.}
	\label{fig:metal_to_dust}
\end{figure*}

\firstround{A different way of visualising what fraction of metals are locked up in dust relative to gas, is to show dust-to-metal ratios: the ratio of the dust mass and the total metal-mass in the galaxy. Since dust consists exclusively of metals, the ratio of the total metal mass and the dust mass is a measure of how much of the metal content in the ISM is locked up in dust grains.} To estimate these, we use the prescription by \citet{Vis2019}:
\begin{equation}
M_Z \equiv f_Z \times M_g + M_d,
\end{equation}
where $M_g$ is the total gas mass, $M_d$ the dust mass, and $f_Z$ is the fraction of metals by mass calculated using
\begin{equation}
f_Z = 27.36 \times \text{O/H}.
\end{equation}
The factor 27.36 comes from the assumption that the Solar metallicity is 12 + log(O/H) = 8.69, and the Solar metal mass fraction Z = 0.0134 \citep{Asplund2009}. \firstround{Using this method we should keep in mind that, in reality, oxygen abundance does not scale directly to the total metal mass, as the oxygen-to-total metal ratio can vary depending on star formation history and stellar metallicity. Similarly, the fraction of oxygen locked up in dust grains does not scale directly with the amount of metals locked up into them, as this depends on the dust composition. Keeping in mind these caveats, the} result is shown in Figure \ref{fig:metal_to_dust}; in Figure \ref{subfig:metalmass_SM} as a function of stellar mass, and in Figure \ref{subfig:metalmass_metal} as a function of metallicity. Markers and symbols are the same as in previous figures. In both panels, dust-to-metal ratios in the Fornax cluster are \firstround{significantly increased compared to those in the field. This implies that a relatively large fraction of the metals in Fornax galaxies are locked up in dust. A similar, though less significant, difference is seen in Virgo galaxies inside $R_\text{vir}$. There is no significant difference between Virgo galaxies outside $R_\text{vir}$ and the DustPedia field sample.}

\section{Resolved gas-to-dust ratios in NGC1436}
\label{sub:NGC1436}
\begin{figure*}
	\centering
	\includegraphics[width=0.7\textwidth]{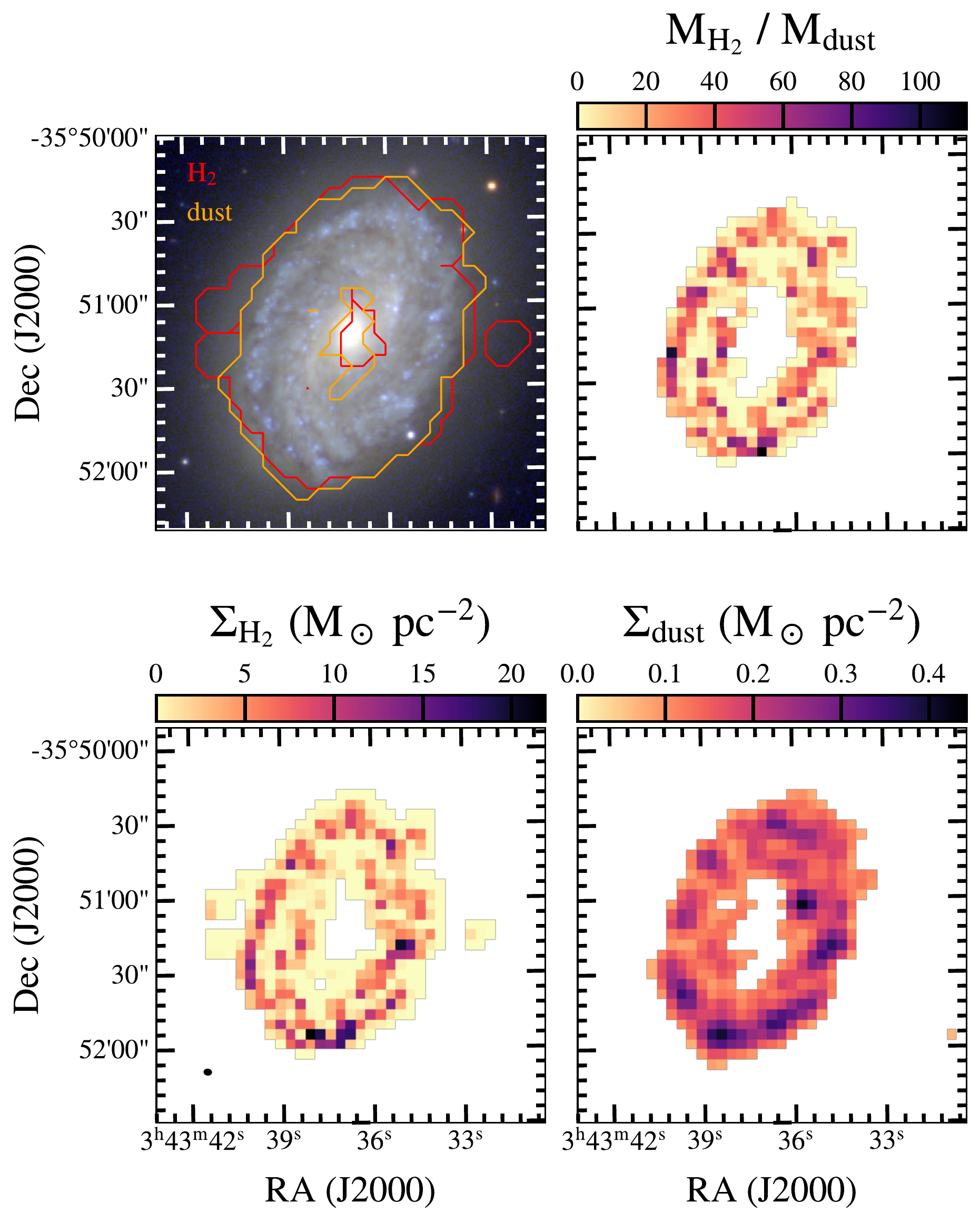}
	\caption{Resolved H$_2$ and dust properties in the flocculent spiral NGC1436. \textit{Upper left panel:} optical \textit{ugb}-image with the extent of the H$_2$ (red) and dust (orange) emission overplotted. \textit{Upper right panel:} resolved H$_2$-to-dust ratios. \textit{Lower-left panel:} H$_2$ surface density at the resolution of the PACS 100 $\mu$m emission. \textit{Bottom-right panel}: dust surface density at the PACS 100 $\mu$m resolution from \textsc{ppmap}. \secround{The beam of the CO observations is shown in the bottom-left corner of the bottom-left panel. There is no clear spatial trend in the H$_2$-to-dust ratios in this galaxy.}}
	\label{fig:NGC1436}
\end{figure*}
Any (radial) trend in the gas-to-dust ratio could give us a hint as to what process might be responsible for the low gas-to-dust (total, H$_2$, and \HI) ratios observed (e.g. the outside-in removal of dust). To investigate whether there might be any radial H$_2$-to-dust gradients \firstround{present in these Fornax galaxies with low gas-to-dust ratios}, and whether the H$_2$-to-dust ratio differs between star forming and more passive regions, we study the spatially resolved H$_2$-to-dust ratio in NGC1436. \secround{While it would be preferred to study resolved total gas-to-dust ratios, this is not feasible, as the ATCA beam is $>$20 times larger than the ALMA beam (and a similar size to the entire molecular disc of this object). \refrep{This will be done in a future paper (Loni et al., in prep.), after completion of the MeerKAT Fornax Survey \citep{Serra2016}}.}

NGC1436 is an almost face-on, flocculent spiral (Figure \ref{fig:NGC1436}, see also Z19 and Z20), making it an ideal candidate to study resolved properties of cluster galaxies in more detail, \firstround{and currently the only galaxy in our sample for which this is possible}. NGC1436 is highly \HI -deficient, and has a high H$_2$-to-\HI\ ratio \citep{Loni2021}. \secround{It has been suggested that this galaxy is undergoing a morphological transition into a lenticular, based on the absence of a clear spiral structure in its outer regions \citep{Raj2019}.} From Figures \ref{fig:total_GD_SM} through \ref{fig:GD_SM_HI}, \refrep{where NGC1436 is highlighted with a filled diamond symbol}, we can see that it has a decreased gas-to-dust ratio, and that this is partly driven by a decreased H$_2$-to-dust ratio, in addition to a decreased \HI -to-dust ratio. 

To ensure we have as complete as possible information on the CO emission, we combine our AlFoCS data with deep observations from the ALMA archive (project ID: 2017.1.00129.S, PI: Kana Morokuma). As part of this program NGC1436 was observed on 30 November 2017 using the Morita array \secround{(see \citealt{Morokuma-Matsui2019} for a description of the data for NGC1316 from this survey).} Its primary beam size is $\sim$ 90$^{\prime \prime}$ at $\sim$115 GHz, and the total area covered is $\sim$182 square arcseconds. The spectral window covering the $^{12}$CO(1-0) was centred on 114.756 GHz, with a bandwidth of 1.875 GHz, covering 3840 channels. The spectral resolution is 5.06 \kms.

\changes{We reduced this data using the \textsc{casa} pipeline (version 5.4.0-68, \citealt{McMullin2007}) before combining it with our own data using the task \textsc{concat}. We then imaged the resulting data set by cleaning it interactively, using the task \textsc{tclean} \citep{Hogbom1974}. A Briggs weighting scheme was used \citep{Briggs1995} with a robust parameter of 0.5. The pixel size in the final datacube is 0\farc 5, and the velocity resolution is 10 \kms. The synthesised beam size is $\sim$2.68 $\times$ 2.06\arcsec. The sensitivity reached is $\sim$2.046 mJy/beam.}

Resolved dust surface-densities are estimated using \textsc{ppmap}. \textsc{ppmap} \citep{Marsh2015} is a Bayesian point-process fitting algorithm that produces image cubes of differential column-density as a function of dust temperature and position. Unlike traditional SED fitting techniques which require all images to be convolved to match the angular resolution of the lowest resolution image, \textsc{ppmap} uses all images in their native resolution (PSFs of each image are also taken as inputs). In our case, this means that we can estimate and plot dust column densities in NGC1436 at the PACS 100 $\mu$m resolution (FWHM $\approx$10$^{\prime \prime}$), the highest resolution of the range of \textit{Herschel} images used to estimate its dust properties.

\refrep{The resulting images are shown in Figures \ref{fig:NGC1436} and \ref{fig:rad_prof}. In the upper-left panel of Figure \ref{fig:NGC1436} the extent of the CO (red) and dust (orange) emission is shown (both are clipped to the 3$\sigma$ level), overplotted on an optical \textit{ugb}-image from the FDS. In the upper-right panel H$_2$-to-dust ratios within the galaxy are shown. The bottom-left and bottom-right panels show the H$_2$ and dust surface densities, respectively. Figure \ref{fig:rad_prof} shows the radial profile of the H$_2$-to-dust ratio, corresponding to the top-right panel in Figure \ref{fig:NGC1436}. Each marker corresponds to the average H$_2$-to-dust ratio in an elliptical annulus (the sum of the ratio in each pixel divided by the total area of the annulus) at each radius, where the radius corresponds to the semi-major axis of the annulus.}

\refrep{There are small-scale variations in the H$_2$-to-dust ratio. Peaks in the H$_2$-to-dust ratio appear to mainly correlate with peaks in the H$_2$ surface density. There is an increase in H$_2$-to-dust ratio with radius, after which it drops off sharply at $R \approx 3.5$ kpc. This means that the outside-in stripping of the gas/dust disc could result in lower integrated H$_2$-to-dust ratios. It could also mean that dust is more easily removed from the outer parts of the disc than molecular gas, causing the gradient observed.}

\refrep{The average H$_2$-to-dust ratio does not exceed $\sim$20 at any radius. This suggests that molecular gas, and likely also dust, can be affected by environment even in the inner parts of galaxies. A comparison with the star formation dominated H$\alpha$ map from MUSE (from F3D, see \S \ref{sub:optical_spectra_fornax}) shows no clear correlation between star forming regions and variations in the H$_2$-to-dust ratio.}


\section{Discussion}
\label{sec:discussion}

\subsection{Low gas-to-dust ratios in the Fornax cluster}
\label{subsec:low_gd}
\firstround{In \S \ref{sub:GD_SM} we have shown that gas-to-dust ratios in Fornax galaxies are \secround{suppressed} compared to a field comparison sample at fixed stellar mass and metallicity.} Decreased total gas-to-dust ratios might be expected in clusters as a result of stripping and truncation of \HI\ discs, which \changes{typically} have scale lengths much larger than H$_2$. Indeed, this is observed both in Fornax cluster galaxies and Virgo galaxies (Figures \ref{fig:total_GD_SM} and \ref{fig:GD_metal_total}). The low gas-to-dust ratios in the Fornax cluster are partly driven by decreased \HI -to-dust ratios compared to field galaxies at fixed mass (Figures \ref{fig:GD_SM_HI} and \ref{fig:GD_metal_HI}). However, these low \HI -to-dust ratios are not the full story. H$_2$-to-dust ratios are also significantly decreased in the Fornax cluster (Figures \ref{fig:GD_SM} and \ref{fig:GD_metal}). In Figure \ref{fig:dust_gas_stellar_mass} we can see that, while dust fractions are decreased in the Fornax cluster compared to field galaxies at fixed stellar mass (panel a), molecular gas fractions are even more strongly decreased (panel b). Broadly, there are three ways in which we could end up with the H$_2$-to-dust ratios observed: 
\begin{enumerate}
\item H$_2$ is destroyed/removed more efficiently than dust,
\item both dust and H$_2$ are destroyed, but the dust reservoir is replenished more efficiently than the H$_2$ reservoir,
\item the physics of the ISM in these cluster galaxies is \secround{unusual} such that ``standard'' observational probes fail to return accurate H$_2$-to-dust ratios.
\end{enumerate}

\subsubsection{Option (i): H$_2$ is destroyed/removed more efficiently than dust}
\firstround{Molecular gas could be destroyed/removed more efficiently than dust by the cluster environment if our sample galaxies had strong radial H$_2$-to-dust gradients, and their gas discs were truncated from the outside in (i.e. by ram pressure stripping). \secround{If this is the case, we are seeing the ``relic'' of a gas/dust reservoir that was larger before the galaxies fell into the cluster.} This could alter the total H$_2$-to-dust mass ratio we would measure. \secround{For example, \citet{Bekki2014} shows that ram pressure stripping can lead to more centrally concentrated star formation.} Studies of nearby galaxies and galaxies in the Virgo cluster are inconclusive as to whether such radial molecular gas-to-dust ratios are observed. Several studies suggest that observed molecular gas-to-dust gradients are driven by a metallicity gradient (i.e. a radial change in \xco, \citealt{Bendo2010, Magrini2011, Pappalardo2012}). Therefore, it is unclear whether an actual (non-\xco -driven) gradient in gas-to-dust ratio is also present.}
 \begin{figure}
	\centering
	\includegraphics[width=0.47\textwidth]{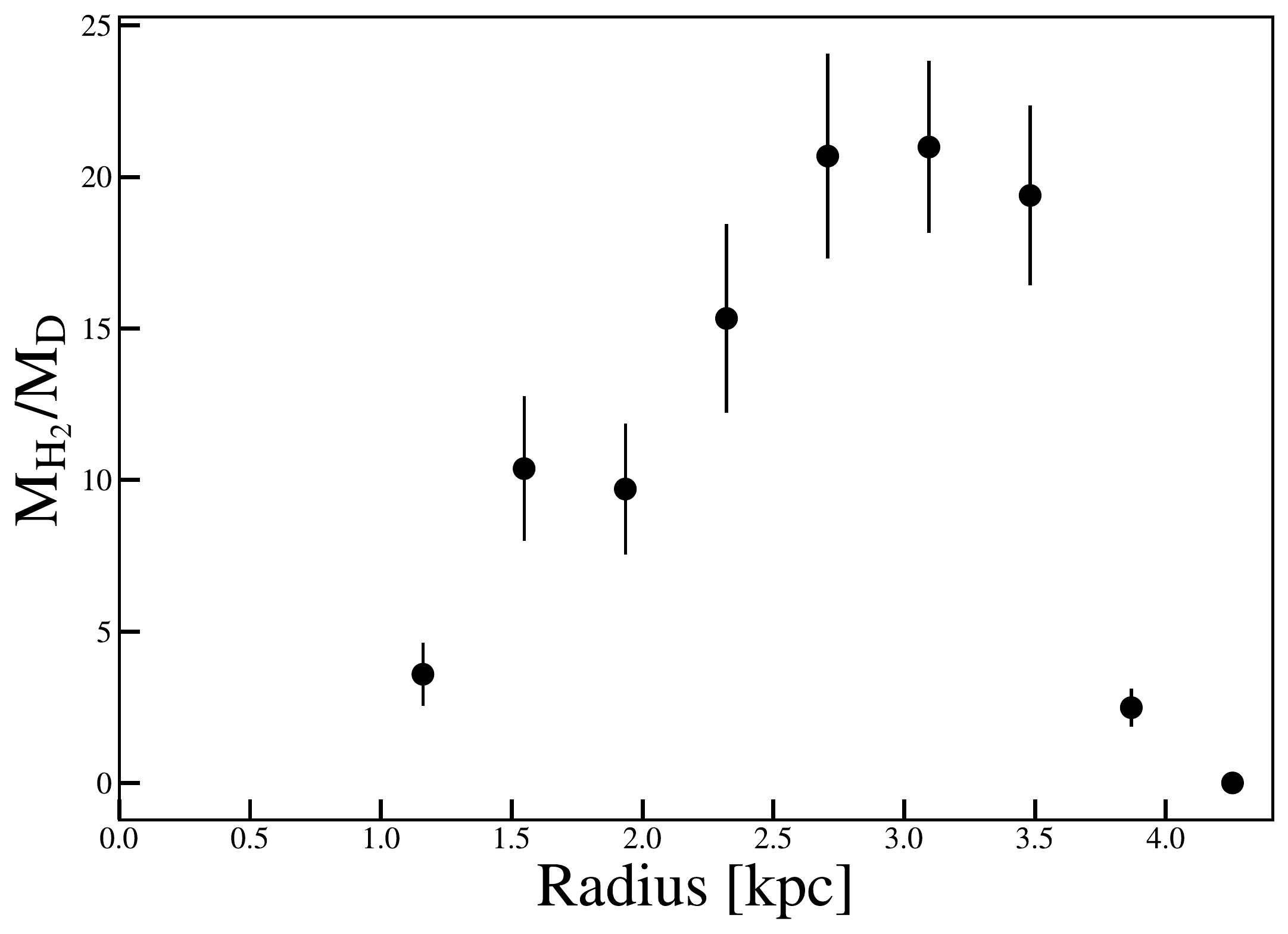}
	\caption{\refrep{Radial profile of the H$_2$-to-dust ratio in NGC1436. It increases with radius, and drops off sharply at $R \approx 3.5$ kpc. The observed gradient implies that outside-in stripping of the gas/dust disc could result in lower integrated H$_2$-to-dust ratios. It could also imply that dust is more easily removed from the outer parts of the disc than molecular gas.}}
	\label{fig:rad_prof}
\end{figure}
\citet{Cortese2010} find that the dust discs of \HI -deficient galaxies are truncated as well as the \HI\ discs. \citet{Corbelli2012} find that gas-to-dust ratios decrease as \HI -deficiency increases, but only up to a certain deficiency threshold, after which they remain constant. \refrep{\HI -deficiency is defined as the difference between the observed \HI\ mass and that expected in an isolated galaxy.} This is because in highly disturbed galaxies both gas and dust are stripped from the inner parts of the galaxies. However, they also find stronger dust than H$_2$ deficiencies in these galaxies. This suggests that, although \HI -deficient galaxies have lower gas-to-dust ratios, and both H$_2$ and dust can be stripped from their inner parts, H$_2$-to-dust ratios in these galaxies remain constant, or even increase slightly. \firstround{This is also observed by \citet{Cortese2016} and in Virgo galaxies in this work, and is discussed further in \S \ref{sub:comparison_virgo}. Since truncation of the gas/dust disc is concluded to result in an \textit{in}crease in H$_2$-to-dust ratios in these two studies, which is also seen here in the Virgo cluster, it is unlikely to be the explanation for the H$_2$-to-dust ratios observed in Fornax.} 
\secround{Observations of the molecular gas and dust in FCC167 (NGC1380, included in this sample) show a nested ISM, in which the dust extends further out than the molecular gas \citep{Viaene2019}. This could indeed mean that the molecular gas in this galaxy was stripped more severely than the dust. However, the molecular gas disc in this galaxy is extremely truncated (e.g. Z19, \citealt{Viaene2019}), and it is unclear how representative this is for the rest of the sample.} \refrep{The observed radial increase in H$_2$-to-dust ratio in NGC1436 (see Figure \ref{fig:rad_prof}) suggests that the stripping of the outer parts of the disc could indeed result in lower integrated H$_2$-to-dust ratios.}

\begin{figure}
	\centering
	\includegraphics[height=0.43\textwidth, width=0.47\textwidth]{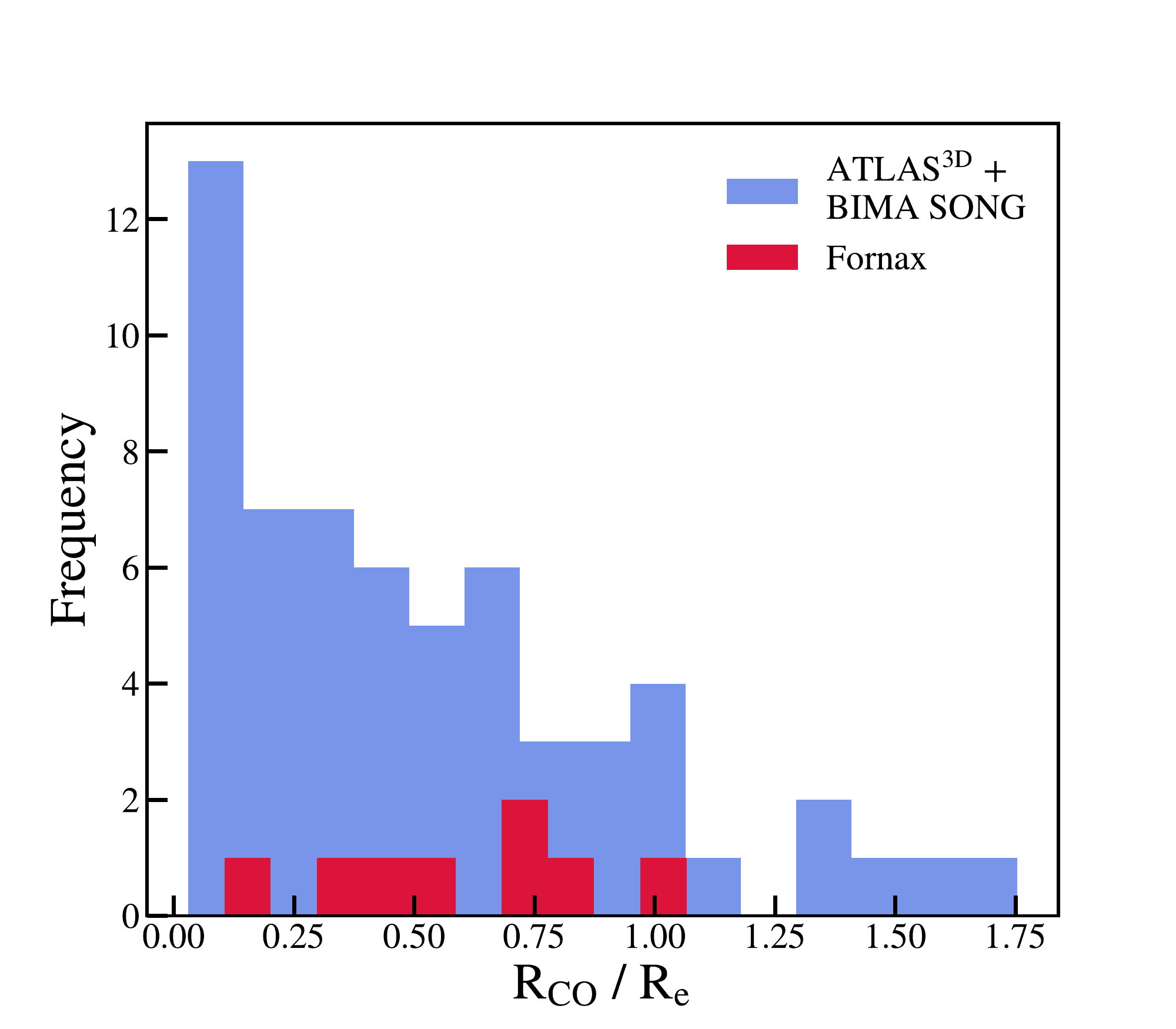}
	\caption{\changes{Histogram of R$_\text{CO}$/Re in the Fornax cluster (crimson) compared to a field sample consisting of gas-rich early-type galaxies from ATLAS$^{\text{3D}}$ (\citealt{Davis2013}) and spiral galaxies from BIMA SONG \citep{Regan2001}, shown in lilac. Fornax cluster galaxies do not have more significantly truncated CO than the field galaxies (a KS test is not able to reject the null-hypothesis that the Fornax and ATLAS$^{\text{3D}}$/BIMA SONG samples are drawn from the same distribution, D=0.29, p=0.5 $\sim 0.67 \sigma$).}}
	\label{fig:gas_extent}
\end{figure}

Figure \ref{fig:gas_extent} shows a histogram of the ratio of R$_\text{CO}$ and effective radius \refrep{R$_e$} (as estimated from the FDS, \citealt{Peletier2020, Venhola2018, Iodice2019a, Raj2019}) in the Fornax cluster (crimson) compared to this ratio in a field sample of gas-rich early-type galaxies (ETGs) from ATLAS$^{\text{3D}}$ \citep{Davis2013} and nearby spiral galaxies from  the Berkeley-Illinois-Maryland Association Survey of Nearby Galaxies (BIMA SONG, \citealt{Regan2001}) in lilac. Although the numbers are small, there is no evidence that Fornax cluster galaxies have smaller R$_\text{CO}$/Re than the field sample (a KS test is not able to reject the null-hypothesis that the Fornax and ATLAS$^{\text{3D}}$/BIMA SONG samples are drawn from the same distribution, D=0.29, p=0.5 $\sim 0.67 \sigma$). 

In order to more generally test whether truncation could play a significant role in creating low H$_2$-to-dust ratios, we create a toy model consisting of an exponential gas/dust disc, with a molecular gas-to-dust ratio that increases linearly outward. We then explore the parameter space describing the shape of the exponential disc and the molecular gas-to-dust gradient resulting in the observed H$_2$ deficiencies (see table 3 in Z19) and molecular gas-to-dust ratios, using a Markov chain Monte Carlo approach. We fix H$_2$ deficiencies and molecular gas-to-dust ratios at the more conservative observed values of 1 and 200, respectively, before truncation, and -1 and 100, respectively, after truncation (see Table \ref{tab:gas-to-dust_ratios} and Table 3 in Z19). \changes{Truncation is simulated by removing a percentage of the disc from the outside in. This percentage is a free parameter.} Although, inevitably, there is a high degree of degeneracy in the parameters describing the exponential disc, it is clear that we would need very steep H$_2$-to-dust gradients and severe truncation ($>$80\% of the gas/dust disc) for this to explain the range of H$_2$-to-dust mass ratios and H$_2$-deficiencies observed. Therefore, \firstround{based on this toy model,} truncation in combination with a H$_2$-to-dust gradient is unlikely to be the sole explanation of the decreased H$_2$-to-dust ratios observed. \firstround{In reality, the spatial extent of the dust may exceed that of the molecular gas, as, for example, observed the nearby spiral galaxy NGC2403 \citep{Bendo2010}. This implies that it would be removed by ram pressure stripping before the molecular gas, as suggested by \citet{Cortese2016}. This implies that it would be even more difficult to remove significant amounts of molecular gas compared to dust in this way.}

\subsubsection{Option (ii): the dust reservoir is replenished more efficiently than the H$_2$ reservoir}
If dust is created at a higher rate than H$_2$ in the star formation cycle, this could result in decreased gas-to-dust ratios. Both gas and dust are observed in supernova remnants (see \citealt{Matsuura2017b} for a review). Recent, in-depth studies of supernova remnants have shown that \changes{large} amounts of dust can be produced in supernovae \citep{Dunne2003, Looze2017, Looze2019, Cigan2019, Priestley2019}. Gas, on the other hand, is produced in relatively small quantities, resulting in gas-to-dust ratios in supernova remnants that are much smaller (5 - 10 times) than typical ISM values \citep{Owen2015, Matsuura2017a, Arias2018, Priestley2019}. Therefore, ongoing star formation in combination with starvation could be (co-)responsible for the decrease in H$_2$-to-dust ratios observed, possibly sped up by the active removal of gas and dust. Several galaxies in the Fornax cluster have slightly increased star formation efficiencies. \secround{Furthermore, starbursts can already have occurred in the pre-processing phase, as galaxies start to interact with the intracluster medium, or in infalling groups \citep{Pinna2019a, Pinna2019b}, although this has mainly been observed in early-type galaxies}. While this does not mean their SFRs or sSFRs are also increased (in fact, the vast majority of them lie below the star formation main sequence, see Z20, so we do not expect them to produce more dust than ``average'' galaxies), it is the H$_2$-to-dust ratio we are interested in. This can be increased as a result of this preferential dust production. 

An inspection of gas-to-dust ratios in the Fornax cluster as a function of depletion time shows that they decrease with increasing depletion time: galaxies currently undergoing starbursts, often showing disturbed molecular gas reservoirs, still have relatively high gas-to-dust ratios. This could mean two things: either this is not a good explanation, or it takes a while for the dust to accumulate, and the effect is only noticeable after the star burst phase is over. \firstround{The latter explanation is supported by the fact that old stellar populations are also a significant source of interstellar dust \citep[while their gas feedback is insufficient to keep star formation going,][]{Matsuura2009, Boyer2012, Hofner2018}. Thus, as gas and dust are stripped and depleted, but star formation has not yet been quenched (the galaxy is experiencing ``starvation'', \citealt{Larson1980}) the dust produced by older stars and supernovae accumulates within the galaxy.} \firstround{Even without ongoing star formation, the dust mass can still continue to increase through interstellar grain growth (assuming some residual cool gas is present in the galaxy to shield the dust from sputtering in the hot intracluster medium, \citealt{Hirashita2012, Mattsson2014, Zhukovska2014}; Galliano et al., submitted). This suggests that the gas-to-dust ratio could even continue to decrease after the star formation has been quenched.}

Many galaxies in the Fornax cluster are deficient in \HI, or even completely devoid of it \firstround{(at the survey sensitivity limit of $\sim 2 \times 10^7 M_\odot$, \citealt{Loni2021}).} If the main formation of H$_2$ is through the condensation of \HI, this could mean that the production of H$_2$ is slower than in non-\HI\ deficient galaxies. However, there also is a significant fraction of \HI -deficient galaxies in the Virgo cluster \citep{Yoon2017}, so if this was the main explanation, we might expect to see similarly low molecular gas-to-dust ratios in the Virgo cluster. \firstround{Moreover, most of these galaxies have high rather than low H$_2$-to-\HI\ ratios, which does not support this theory \citep{Loni2021}.}

\subsubsection{Option (iii): altered ISM physics result in inaccurate gas-to-dust ratios}
It is possible that the radiation field in the cluster destroys CO (but not H$_2$), leading us to underestimate H$_2$-to-dust ratios. In the Galaxy, the radiation field only affects the outermost layers of molecular clouds (e.g. \citealt{Maloney1988}). However, if the radiation field is more intense, it can affect the entire cloud. In fact, besides metallicity, the strength of the FUV radiation field is the most important factor determining \xco\ e.g. \citealt{Bisbas2015} and references therein). It is possible that X-rays from the hot intracluster medium (e.g. \citealt{Jones1997}) have a similar effect. However, it would be difficult to destroy CO while leaving dust grains intact. Moreover, in that case we might expect to see a similar or stronger decrease in H$_2$-to-dust ratios in the more massive Virgo cluster, which is not observed.

An alternative factor that could be affecting our measurements is the dust composition. Large silicate grains are not as easily destroyed by the radiation field, which means that H$_2$ might be destroyed while leaving a large fraction of the dust unaffected. \secround{Indeed, in their detailed analysis of the dust content of FCC167, \citet{Viaene2019} find that almost no small grains are present in the dust reservoir of this galaxy. Only larger, self-shielding grains survive the effects of the cluster environment.} Another possibility is that larger grains are \firstround{broken up} by the radiation field, resulting in a relatively high fraction of small dust grains. If we then assume a standard dust composition with a lower fraction of small grains to estimate the dust mass, we could be overestimating the dust mass. \firstround{It seems, however, more likely that such a radiation field would continue to destroy bonds until no significant amount of dust is left.} \secround{Moreover, this contradicts the observations of FCC167 described above.}

\firstround{Furthermore, the carbon-to-silicate ratio of the dust grains can play a role. \secround{Due to environmental conditions and typical grain sizes, silicate grains can have longer lifetimes ($\sim 1$ Gyr) than carbon grains (of the order 100 Myr, e.g. \citealt{Slavin2015})}. If the composition of dust grains in the Fornax cluster is relatively silicate-rich, more oxygen may be expected to be locked up in these grains.} \\

\firstround{Low gas-to-dust ratios have also been observed in field galaxies, for example in the early-type galaxy NGC5485. This object has a total G/D of $<$14.5, as estimated from FIR, CO, and \HI\ observations \secround{(\citealt{Baes2014}, the upper limit is the result of non-detection of \HI\ and H$_2$ in sensitive observations obtained \refrep{as part of} ATLAS$^{3\text{D}}$)}. Suggested explanations for this observation include the majority of the gas being warm or hot rather than cold, the presence of a relatively large fraction of CO-dark gas, or a recent merger with a metal-poor dwarf galaxy. There could be a larger fraction of warm/hot gas present in Fornax galaxies if the gas at larger scale-heights is affected by the surrounding hot intracluster medium. However, the molecular gas is dense and centrally located, so it would be difficult to explain the decreased H$_2$-to-dust ratios in this way. Furthermore, there is no obvious reason why Fornax galaxies would contain more CO-dark gas than field galaxies (at fixed stellar mass), as their metallicities are average or even high compared to the DustPedia field sample (see \S \ref{sub:GD_SM}). Moreover, the latter explanation is unlikely to apply to the entire Fornax sample. Therefore, it seems more likely that the observed low gas-to-dust ratios are a result of environmental effects.} 

\subsection{Comparison with the Virgo cluster}
\label{sub:comparison_virgo}
\firstround{In this work, the main difference between the Fornax and Virgo clusters is that the low total gas-to-dust ratios in Fornax are partly driven by decreased H$_2$-to-dust ratios, while these are \textit{in}creased inside the virial radius of the Virgo cluster. If Virgo galaxies within twice its virial radius are included, this result disappears (see Figures \ref{fig:total_GD_SM} - \ref{fig:GD_metal_HI} and Table \ref{tab:KS_results}). Thus, in Virgo, decreased total-gas-to-dust ratios are purely driven by the low \HI -to-dust ratios. This is in agreement with \citet{Cortese2016}, who find that the decrease in total gas-to-dust ratios in \HI -poor galaxies is purely driven by the decrease in \HI -to-dust ratios, while H$_2$-to-dust ratios are increased compared to field galaxies at fixed mass and metallicity. They attribute these anomalous ratios to the different spatial distribution of the two gas phases and the dust, resulting in the differential stripping of the ISM. \secround{As described above, observations of the ISM in FCC167 indeed show a nested ISM, in which the the molecular gas disc is more compact than the dust ring present in this galaxy \citep{Viaene2019}. However, since this galaxy is significantly deficient in molecular gas, it is unclear whether this nested ISM was already present before the galaxy entered the cluster, or if this is the result of the more efficient stripping of molecular gas compared to dust by the cluster environment.}} 

\firstround{The two main differences between the Fornax and Virgo cluster are that Virgo is more dynamically active, while Fornax is more dynamically evolved, and the Virgo cluster is much more massive than the Fornax cluster (see \S \ref{sec:introduction}). The former difference could be an explanation for option ii: dust is accumulated in the star formation process while H$_2$ is slowly depleted. It could be that there is an initial boost in H$_2$ content, after which it stops being replenished, and this effect becomes more visible over time. However, since the galaxies studied here all still have a detectable ISM, which is not expected after several pericentric passages, it is unlikely that the Fornax galaxies have spent significantly more time in the cluster than the Virgo galaxies, or they would have no gas left.}

\firstround{It may be possible that ram pressure stripping results in the increased H$_2$-to-dust ratios observed in the Virgo cluster and in the galaxies studied by \citet{Cortese2016} if dust is stripped before molecular gas. This is supported by an ongoing study of the Virgo cluster (Zabel et al., in prep.), where we observe normal to high H$_2$ masses in the Virgo galaxies most affected by (past and ongoing) ram pressure stripping. It is possible that ram pressure compresses atomic gas, aiding its condensation into H$_2$, and thus contributing to the increase in H$_2$-to-dust ratios. \secround{This has indeed been observed in jellyfish galaxies \citep{Moretti2020}.} Since the Fornax cluster is less massive than Virgo, ram pressure stripping might play less of a role there, or not be strong/violent enough to quickly remove \HI\ or compress it into H$_2$. It is possible that starvation is more important in Fornax, slowly depriving its galaxies of H$_2$, while the dust continues to accumulate. However, this is mostly speculation, and the wildly different results between both clusters are difficult to explain. Expansion of this sort of analysis to other clusters is clearly needed to determine the true cause of this effect.}

\subsection{Gas-to-dust vs. metallicity}
\label{subsec:gd_vs_metal}
It is often assumed that the dust-to-metal ratio is constant with time, due to the dust formation timescale being similar to the dust destruction timescale. A constant dust-to-metal ratio implies that the gas-to-dust ratio depends on metallicity as G/D $\propto$ Z$^{-1}$, referred to as the ``reference trend'' by \citet{RemyRuyer2014}. This trend has been shown to hold for galaxies with metallicities close to solar, but to break down at the low-metallicity regime where dwarf galaxies are found, which have higher observed gas-to-dust ratios than predicted by this trend (\citealt{RemyRuyer2014} and references therein). \firstround{However, a recent review by \citet{Peroux2020} has shown that the relation between metallicity and dust-to-metal ratio indeed holds in the low-metallicity/high redshift regime, albeit with more scatter due to the more complex chemistry here.} From Figures \ref{fig:GD_metal_total} and \ref{fig:metal_to_dust} we can see that gas-to-dust ratios in the galaxies in our sample are consistently lower than those in the DustPedia sample, suggesting that they do not follow such a reference trend. This implies that their chemical evolution is different from regular field galaxies, which likely has to do with the environment they reside in. Alternatively, the physics of the ISM, \secround{such as the radiation field or dust composition,} in these galaxies could be different such that they contain unusually high fractions of ``CO-free'' \firstround{molecular} gas, as discussed above.

Figures \ref{fig:GD_metal_total} through \ref{fig:metal_to_dust} imply that the fraction of metals locked up in dust (versus gas) is high in the Fornax cluster. If this is the result of stripping of high-metallicity gas (and dust), these are enriching the intracluster medium. \firstround{Some Fornax galaxies are quite metal-rich, while they still have significant amounts of gas left. This is in broad agreement with what is found by \citet{Hughes2013}, who find that gas-poor galaxies are relatively metal-rich. This could be due to the more metal-poor outer regions of the galaxy being stripped first. However, they do not find any significant difference between metallicities in the Virgo cluster and in the field. These metal- and gas-rich galaxies could indicate that we are observing a special time in the lifespan of the galaxies (i.e. we expect many of them to be on their first infall, \secround{which is also suggested by \citealt{Loni2021}}), or witnessing a special time for the cluster itself.}

\section{Summary}
\label{sec:summary}
We have studied gas-to-dust ratios in a sample of 15 Fornax galaxies from the ALMA Fornax Cluster Survey (AlFoCS) as a function of stellar mass. In addition, in a sub-sample of 9 galaxies that were also observed with VLT/MUSE, as part of the Fornax3D project, the gas-to-dust ratio was also studied as a function of metallicity. We have separated H$_2$ and \HI\ to separately study H$_2$-to-dust and \HI -to-dust ratios. \firstround{Each gas-to-dust ratio (from \HI, H$_2$, and \HI\ + H$_2$) was compared to a field sample and galaxies in the Virgo cluster (separated into galaxies inside $R_\text{vir}$ and inside 2$R_\text{vir}$), both from DustPedia. Dust, H$_2$, \HI, and stellar masses were calculated using the same assumptions and methods as DustPedia where possible, to maximise homogeneity between both samples.} We also studied dust-to-metal ratios as a function of stellar mass and metallicity, and H$_2$-to-dust ratios as a function of distance from the cluster centre. We made use of \textsc{ppmap} to study resolved H$_2$-to-dust ratios in NGC1436, an almost face-on, flocculent spiral galaxy, at the PACS 100 $\mu$m resolution. Our main conclusions are as follows:
\begin{itemize}
\item Gas-to-dust ratios in the Fornax cluster are systematically decreased compared to the field. \refrep{Kolmogorov-Smirnov (KS) and Anderson-Darling (AD) tests are able to reject the hypothesis that the Fornax sample and DustPedia field sample are drawn from the same distribution at $\gg 5 \sigma$. According to the same tests, this difference is not only driven by \HI\ deficiencies, but H$_2$-to-dust ratios are also decreased. We propose a number of explanations for \changes{this}:}
\begin{itemize}
\item[-] H$_2$ is destroyed\changes{/removed from the galaxies} \firstround{more efficiently than dust}. This is possible if there is a radial H$_2$-to-dust gradient in combination with a truncated H$_2$/dust disc. \firstround{However, Fornax galaxies do not show any evidence of having significantly truncated H$_2$ discs. Moreover, past studies of \HI -deficient galaxies and galaxies in the Virgo cluster suggest that dust is stripped before molecular gas.} Finally, from inspection of a toy model, one would need quite steep radial gradients, in combination with severe truncation \changes{of the gas disc in these objects}. Thus, it seems unlikely that this is the sole explanation for the low ratios observed. \refrep{However, the radial increase in H$_2$-to-dust ratio observed in NGC1436 suggests that this effect can contribute.}
\item[-] Both H$_2$ and dust are destroyed, but the dust reservoir is replenished more efficiently than the H$_2$ reservoir. This is possible if relatively large amounts of dust are created by the deaths of massive stars, \firstround{and/or H$_2$ is not replenished as efficiently as a result of starvation, stripping of \HI, and/or the inefficient condensation of \HI\ into H$_2$. Recent observations of gas-to-dust ratios in supernova remnants, as well as the known production of dust by old stellar populations, and interstellar grain growth, suggest that this could be a possible explanation. The slightly increased star formation efficiencies observed in Z19 possibly speed up this process. However, the H$_2$-to-\HI\ ratios in Fornax galaxies are high compared to those in field galaxies at fixed stellar mass, suggesting that the conversion of \HI\ into H$_2$ does take place efficiently \citep{Loni2021}.}
\item[-] The physics of the ISM is altered in such a way that ``standard'' assumptions we make in order to estimate H$_2$ and dust masses are no longer valid. For example, the strong radiation field in the cluster could disintegrate CO, while leaving H$_2$ intact, leading us to underestimate H$_2$ masses. Alternatively, the dust composition could be altered by the radiation field, possibly resulting in a higher fraction of small grains, which leads us to overestimate the dust mass. We consider these explanations less likely, however we cannot rule them out.
\end{itemize}

\item Gas-to-dust ratios as a function of metallicity are also decreased, while dust-to-metal ratios are increased. This suggests that a relatively large fraction of metals is locked up in dust in these Fornax galaxies.
\item \refrep{Total gas-to-dust ratios in the Virgo cluster are decreased significantly, especially inside its virial radius. However, unlike in Fornax, this decrease is purely driven by a decrease in \HI -to-dust ratios, while H$_2$-to-dust ratios are \textit{in}creased. The difference in dynamical state and mass between both clusters are suggested as possible explanations for this, however the differences between the two clusters remain puzzling.}
\item Gas-to-dust ratios do not show any obvious variation with (projected) cluster-centric distance, although this is difficult to measure because of projection effects and small-number statistics. \changes{In the Virgo cluster there is possibly a weak correlation between cluster-centric distance and gas-to-dust ratio (a Kendall's Tau test returns a tau statistic of 0.23 with a p-value of 0.04). \firstround{This likely reflects the difference between galaxies inside and outside $R_\text{vir}$, but is diluted by projection effects.}}
\item \refrep{Resolved H$_2$-to-dust ratios in NGC1436 show an increase with ratio, ending in a sharp drop at $R \approx 3.5$ kpc. There are very few pixels that exceed a ratio of $\sim$40, suggesting that, aside from small-scale variations, low H$_2$-to-dust ratios are low throughout the entire gas/dust disc. There is no clear correlation with H$\alpha$ emission.}
\end{itemize}

In summary, total gas-to-dust, \HI -to-dust and H$_2$-to-dust ratios in the Fornax cluster are significantly decreased compared to those in field galaxies at fixed stellar mass and metallicity. Their significantly increased dust-to-metal ratios suggest that a relatively large fraction of metals in these galaxies is locked up in dust. There is a variety of \changes{environmental} mechanisms that could explain this, \secround{and it is possible that two or all three of the suggested explanations play a role}. \firstround{We see an opposite effect inside the virial radius of the Virgo cluster, where decreased total gas-to-dust ratios are purely driven by decreased \HI -to-dust ratios, while H$_2$-to-dust ratios are \textit{in}creased. These results disappear when galaxies inside 2$R_\text{vir}$ are included. There are several differences between both clusters that could possibly explain this, however, it remains a puzzling result.} \\

\firstround{Further study is clearly required to determine what is driving the low gas-to-dust mass ratios in cluster galaxies. To eliminate any \xco\ related effects on the gas-to-dust ratio, one could attempt to estimate \xco\ from the IR emission, simultaneously with dust masses. Such an approach is, for example, taken by \citet{Leroy2011} and \citet{Sandstrom2013}. Their models, however, rely on resolved measurements, which are not available for our sample. If suitable data could be obtained it would interesting to investigate this further in a future work.} \\

\firstround{Furthermore, it is crucial to expand this analysis to study other galaxy clusters and groups. An ongoing, similar study of H$_2$-to-dust ratios in the Coma cluster (Zabel et al., in prep.) will possibly help disentangle which environmental effects and cluster properties have a significant effect on these ratios.}

\section*{Acknowledgements}
We would like to thank the anonymous referee for taking the time to revise and provide constructive feedback on our manuscript.

The authors would also like to thank the (other) dust experts at the PHYSX department at Cardiff University for helpful suggestions, and pointing us to some useful literature.

NZ acknowledges support from the European Research Council (ERC) in the form of Consolidator Grant CosmicDust (ERC-2014-CoG-647939).

TAD acknowledges support from the Science and Technology Facilities Council through grant ST/S00033X/1.

R.F.P. and E.I. acknowledge financial support from the European Union's Horizon 2020 research and innovation program under the Marie Sk\l odowska-Curie grant agreement No. 721463 to the SUNDIAL ITN network.

This project has received funding from the European Research Council (ERC) under the European Union's Horizon 2020 research and innovation programme (grant agreement no. 679627; project name FORNAX).

This work made use of the H$_2$ mass data of DustPedia late-type galaxies \citep{Casasola2020, Davies2017}.

DustPedia is a collaborative focused research project supported by the European Union under the Seventh Framework Programme (2007-2013) call (proposal no. 606847). The participating institutions are: Cardiff University, UK; National Observatory of Athens, Greece; Ghent University, Belgium; Universit\'e Paris Sud, France; National Institute for Astrophysics, Italy and CEA, France.

This paper makes use of the following ALMA data: ADS/JAO.ALMA\#2015.1.00497.S, and ADS/JAO.ALMA\#2017.1.00129.S. ALMA is a partnership of ESO (representing its member states), NSF (USA) and NINS (Japan), together with NRC (Canada), MOST and ASIAA (Taiwan), and KASI (Republic of Korea), in cooperation with the Republic of Chile. The Joint ALMA Observatory is operated by ESO, AUI/NRAO and NAOJ.

The Australia Telescope is funded by the Commonwealth of Australia for operation as a National Facility managed by CSIRO. We acknowledge the Gomeroi people as the traditional owners of the Observatory site.

This research made use of Photutils, an Astropy package for
detection and photometry of astronomical sources \citep{Bradley2019}.

This research made use of Astropy,\footnote{http://www.astropy.org} a community-developed core Python package for Astronomy \citep{Astropy2013, Astropy2018}.

This research made use of APLpy, an open-source plotting package for Python \citep{Robitaille2012}.

\section{Data Availability}
The data underlying this paper are available in the ALMA archive: ADS/JAO.ALMA\#2015.1.00497.S, and ADS/JAO.ALMA\#2017.1.00129.S, the ESO archive: 296.B-5054, and the Australia Telescope Online Archive: C2894. \refrep{The updated \textit{Herschel} maps from the \textit{Herschel} Fornax Cluster Survey are available from \url{http://www.mwlsmith.co.uk/surveys.html}.}




\bibliographystyle{mnras}
\bibliography{References} 


\clearpage

\bsp	
\label{lastpage}
\end{document}